\documentclass{aa}
\usepackage{longtable}
\usepackage{supertabular}
\usepackage{graphics}
\usepackage{rotating}
\usepackage{natbib}
\begin{document}
\thesaurus{0.8(0.9.10.1; 0.9.13.2; 0.8.16.5)}
\title{RASS-SDSS Galaxy Cluster Survey.} 
\subtitle{IV. A ubiquitous dwarf galaxy population in clusters.}
\author{P. Popesso\inst{1}, A. Biviano\inst{2}, H. B\"ohringer\inst{1}, 
M. Romaniello\inst{3}}
\institute{ Max-Planck-Institut fur extraterrestrische Physik, 85748 Garching, 
Germany
\and INAF - Osservatorio Astronomico di Trieste, via G. B. Tiepolo 11, I-34131,
Trieste, Italy
\and European Southern Observatory, Karl Scharzschild Strasse 2, D-85748}
\authorrunning{P. Popesso et al.}
\maketitle

\begin{abstract}
We analyze the Luminosity Functions (LFs) of a subsample of 69
clusters from the RASS-SDSS galaxy cluster catalog. When calculated
within the cluster physical sizes, given by $r_{200}$ or $r_{500}$,
all the cluster LFs appear to have the same shape, well fitted by a
composite of two Schechter functions with a marked upturn and a
steepening at the faint-end. Previously reported cluster-to-cluster
variations of the LF faint-end slope are due to the use of a metric
cluster aperture for computing the LF of clusters of different masses.

We determine the composite LF for early- and late-type galaxies, where
the typing is based on the galaxy $u-r$ colors. The late-type LF is
well fitted by a single Schechter function with a steep slope
($\alpha=-2.0$ in the $r$ band, within $r_{200}$). The early-type LF
instead cannot be fitted by a single Schechter function, and a
composite of two Schechter functions is needed. The faint-end upturn
of the global cluster LF is due to the early-type cluster
galaxies. The shape of the bright-end tail of the early-type LF does
not seem to depend upon the local galaxy density or the distance from
the cluster center. The late-type LF shows a significant variation
only very near the cluster center.  On the other hand, the faint-end
tail of the early-type LF shows a significant and continuous variation
with the environment.

We provide evidence that the process responsible for creating the
excess population of dwarf early type galaxies in clusters is a
threshold process that occurs when the density exceeds $\sim 500$
times the critical density of the Universe.

We interpret our results in the context of the 'harassment' scenario,
where faint early-type cluster galaxies are predicted to be the
descendants of tidally-stripped late-type galaxies.
\end{abstract}

\section{Introduction}
\label{s-intro}
The galaxy Luminosity Function (LF) is a fundamental tool for
understanding galaxy evolution and faint galaxy populations. The shape
of the cluster LF provides information on the initial formation and
subsequent evolution of galaxies in clusters while the slope of the
faint-end indicates how steeply the dwarf number counts rise as a
function of magnitude.

Much work has been done on the cluster LF, with various groups finding
differences in its shape and the faint-end slope.  Different
techniques have been used to measure LFs of individual clusters or to
make a composite LF from individual clusters LFs (e.g. Dressler 1978;
Lugger 1986, 1989; Colless 1989; Biviano et al. 1995; Lumsden et al.
1997; Valotto et al. 1997; Rauzy et al.  1998; Garilli et al.  1999;
Paolillo et al. 2001; Goto et al.  2002; Yagi et al. 2002; Popesso et
al. 2004a). Whether the LF of cluster galaxies is universal or not, and
whether it is different from the LF of field galaxies are still
debated issues.  Several authors (Dressler 1978; Lumdsen et al. 1997;
Valotto et al. 1997; Garilli et al. 1999; Goto et al. 2002; Christlein
\& Zabludoff 2003) have found significant differences between the LFs
of different clusters as well as between the LFs of cluster and field
galaxies, while others (Lugger 1986, 1989; Colless 1989; Rauzy et
al. 1998; Trentham 1998; Paolillo et al. 2001; Andreon 2004) have
concluded that the galaxy LF is universal in all environments. Another
debated issue is the slope of the faint end of the LF of cluster
galaxies (see, e.g., Driver et al. 1994; Lobo et al. 1997; Smith et
al. 1997; Phillipps et al.  1998; Boyce et al.  2001; Beijersbergen et
al.  2001; Trentham et al. 2001; Sabatini et al. 2003; Cortese et
al. 2003). The LF of cluster galaxies is typically observed to steepen
faint-ward of $M_g \sim -18$, with power-law slopes $\alpha \sim -1.8
\pm 0.4$. This corresponds to the debated upturn of the cluster LF due
to an excess of dwarf galaxies relative to the field LF. The effect
may be real, and due to cluster environmental effects, but it could
also be generated by systematics in the detection techniques of faint,
low surface-brightness galaxies.

In Popesso et al. (2004a, hereafter paper II) we have recently
analyzed the LF of clusters from the RASS-SDSS (ROSAT All Sky Survey
-- Sloan Digital Sky Survey) galaxy clusters survey down to $-14$
mag. We concluded that the composite cluster LF is characterized by an
upturn and a clear steepening at faint magnitudes, in all SDSS
photometric bands. Different methods of background subtraction were
shown to lead to the same LF. The observed upturn of the LF at faint
magnitudes was shown in particular not to be due to background
contamination by large scale structures or multiple clusters along the
same line of sight. We concluded that the observed steepening of the
cluster LF is due to the presence of a real population of faint
cluster galaxies.

The composite LF was well fitted by the sum of two Schechter (1976)
functions.  The LF at its bright-end was shown to be characterized by
the classical slope of $-1.25$ in all photometric bands, and a
decreasing $M^*$ from the $z$ to the $g$ band.  The LF at its
faint-end was found to be much steeper than the LF at its bright-end,
and characterized by a power-law slope $-2.5 \le \alpha \le -1.6$.
The observed upturn of the LF was found to occur at $-16$ in
the $g$ band, and at $-18.5$ in the $z$ band.

A steep mass function of galactic halos is a robust prediction of
currently popular hierarchical clustering theories for the formation
and evolution of cosmic structure (e.g. Kauffmann et al.
1993; Cole et al. 1994). This prediction conflicts with the flat
galaxy LF measured in the field and in local groups, but is in
agreement with the steep LF measured in the RASS-SDSS clusters. Two
models have been proposed to explain the observed environmental
dependence of the LF. According to Menci et al. (2002), merging
processes are responsible for the flattening of the LF; the
environmental dependence arises because mergers are more common in the
field (or group) environment than in clusters, where they are
inhibited by the high velocity dispersion of galaxies. According to
Tully et al. (2002), instead, the LF flattening is due to inhibited
star formation in dark matter halos that form late, i.e. after
photoionization of the intergalactic medium has taken place.  Since
dark matter halos form earlier in higher density environments, a
dependence of the observed LF slope on the environment is predicted.
On the other hand, if reionization happens very early in the Universe,
this scenario may not work (Davies et al. 2005).  Other physical
processes are however at work in the cluster environment, such as
ram-pressure stripping (Gunn \& Gott 1972) and galaxy harassment (e.g.
Moore et al. 1996, 1998), which are able to fade cluster galaxies,
particularly the less massive ones. Whether the outcome of these
processes should be a steepening or a flattening of the LF faint-end
is still unclear.

In paper II it was also shown that the bright-end of the LF is
independent from the cluster environment, and the same in all
clusters. On the other hand, the LF faint-end was found to vary from
cluster to cluster. In the present paper (IV in the series of the
RASS-SDSS galaxy cluster survey) we show that the previously found
variations of the faint end of the cluster LF are due to aperture
effects. In other words, when measured within the physical size of the
system, given by either $r_{200}$ or $r_{500}$, the LF is invariant
for all clusters, both at the bright and at the faint end. We also
analyze how the number ratio of dwarf to giant galaxies in galaxy
clusters depends on global cluster properties such as the velocity
dispersion, the mass, and the X-ray and optical luminosities. Finally,
we investigate the nature of the dwarf galaxies in clusters by
studying their color distribution and suggest a possible formation
scenario for this population.

The paper is organized as follows. In \S~\ref{s-data} of the paper we
describe our dataset. In \S~\ref{s-lfmethod} we summarize the methods
used to calculate the individual and the composite cluster LFs. In
\S~\ref{s-glob} we summarize our methods for measuring the clusters
characteristic radii. In \S~\ref{s-lf} we analyze the resulting
composite and individual LFs.  In \S~\ref{s-type} we determine the
cluster composite LF per galaxy type. In \S~\ref{s-envi} we analyse
the environmental dependence of the LF, and compare the cluster and
field LFs.  In \S~\ref{s-disc} we provide our discussion, suggesting a
possible formation scenario for the faint galaxy population in
clusters. Finally, in \S~\ref{s-conc} we draw our conclusions.

For consistency with paper II and with previous works, we use
$\rm{H}_0=100 \; \rm{h} \; \rm{km} \; \rm{s}^{-1} \; \rm{Mpc}^{-1}$,
$\Omega_{m}=0.3$ and $\Omega_{\Lambda}=0.7$ throughout the paper.

\section{The data}
\label{s-data}
In order to study the variation of the cluster LF from system to
system, the analysis has to be applied to a large statistical sample
of clusters, covering the whole spectrum of properties (in mass,
richness, X-ray luminosity and optical luminosity) of the systems
considered. Since the X-ray observations provide a robust method of
identification of galaxy clusters and the X-ray luminosity is a good
estimator of the system total mass and optical luminosity (see paper I
and Popesso et al. 2004c, hereafter paper III), we have used for our
purpose the RASS-SDSS galaxy cluster sample, which is an X-ray
selected sample of objects in a wide range of X-ray luminosity.  The
updated version of the RASS-SDSS galaxy cluster catalog comprises 130
systems detected in the RASS and in the SDSS sky region (16 clusters
more than in the first version of the catalog released in paper I due
to the larger sky area available in the SDSS DR2). The X-ray cluster
properties and the cluster redshifts have been taken from a variety of
X-ray catalogs, that allow to cover the whole $L_X$ spectrum. The
X-ray intermediate and bright clusters have been selected from three
ROSAT based cluster samples: the ROSAT-ESO flux limited X-ray cluster
sample (REFLEX, B\"ohringer et al. 2002), the Northern ROSAT All-sky
cluster sample (NORAS, B\"ohringer et al.  2000), the NORAS 2 cluster
sample (Retzlaff 2001). The X-ray faint clusters and the groups have
been selected from two catalogs of X-ray detected objects: the ASCA
Cluster Catalog (ACC) from Horner (2001) and the Group Sample (GS) of
Mulchaey et al.  (2003). The RASS-SDSS galaxy cluster sample comprises
only nearby systems at the mean redshift of 0.1. The sample covers the
entire range of masses and X-ray luminosities, from very low-mass and
X-ray faint groups ($10^{13} M{\odot}$ and $10^{42} erg s^{-1}$) to
very massive and X-ray bright clusters ($5\times10^{15} M{\odot}$ and
$5\times10^{44} erg s^{-1}$).

The optical photometric data are taken from the 2$^{nd}$ data release
of the SDSS (Fukugita et al.  1996, Gunn et al. 1998, Lupton et
al. 1999, York et al. 2000, Hogg et al. 2001, Eisenstein et al. 2001,
Smith et al. 2002, Strauss et al. 2002, Stoughton et al.  2002,
Blanton et al. 2003 and Abazajian et al.  2003).  The SDSS consists of
an imaging survey of $\pi$ steradians of the northern sky in the five
passbands $u, g, r ,i, z,$. The imaging data are processed with a
photometric pipeline (PHOTO) specially written for the SDSS data.  For
each cluster we defined a photometric galaxy catalog as described in
\S~3 of Popesso et al. (2004b, paper I). For the analysis in this
paper we only use SDSS Model magnitudes (see paper II for details).

In this paper we consider a subsample of 69 clusters of the
RASS-SDSS sample for which the masses, velocity dispersion, $r_{200}$
and $r_{500}$ (see \S~\ref{s-glob}) were derived through the virial
analysis (see paper III) applied to the
spectroscopic galaxy members of each systems. 

Since throughout the paper the results obtained with the current
analysis of the cluster LF are often compared with the results
obtained in paper II, it is important to notice that the cluster
sample used here is a subsample of the dataset used in paper II. 

\begin{figure*}
\begin{center}
\begin{minipage}{0.8\textwidth}
\resizebox{\hsize}{!}{\includegraphics{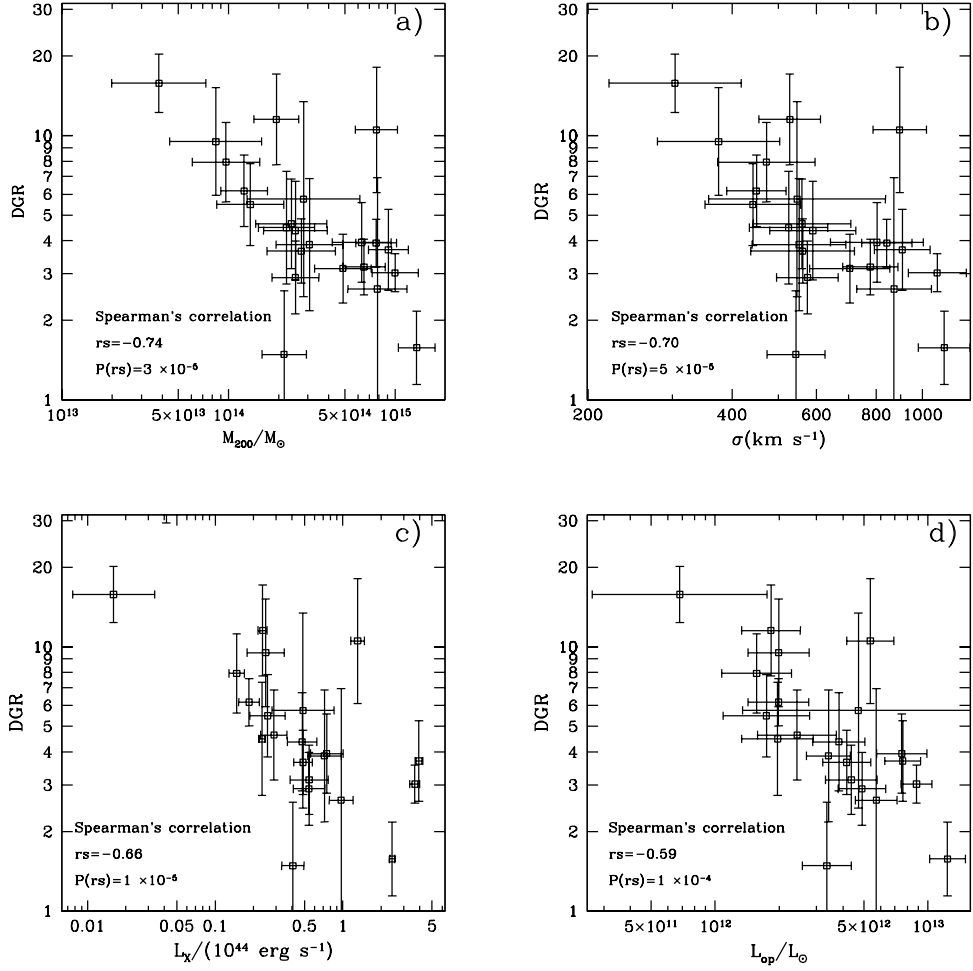}}
\end{minipage}
\end{center}
\caption{The $z$-band DGR vs. the
cluster mass (panel a), the velocity dispersion $\sigma$ (panel
b), the X-ray luminosity (panel $c$), and the optical
luminosity (panel $d$). The DGR is calculated within a circle of 1 Mpc
radius centered on the X-ray cluster center. In each panel, we list
the value of the Spearman's rank correlation coefficient and the
implied probability of no correlation.}
\label{sommario}
\end{figure*}

\section{Determination of the individual and composite Luminosity Functions}
\label{s-lfmethod}
We here summarize the methods by which we measure the individual and
composite cluster LFs. Full details can be found in papers I and II.

We consider two different approaches to the statistical subtraction of
the galaxy background. As a first approach, we calculate a local
background in an annulus centered on the X-ray cluster center with an
inner radius of 3 $\rm{h}^{-1}$ Mpc and a width of 0.5 deg.

As a second approach we derive a global background correction. We
define as $N_{bg}^g(m)dm$ the mean of the galaxy number counts
determined in five different SDSS sky regions, randomly chosen, each
with an area of 30 $\rm{deg^2}$.  A detailed comparison of the local
and global background estimates can be found in paper I. The results
shown in this paper are obtained using a global background
subtraction.

We derive the LFs of each cluster by subtracting from the galaxy
counts measured in the cluster region, the field
counts rescaled to the cluster area. Following previous literature
suggestions, we exclude the brightest cluster galaxies from the
clusters LFs.

In order to convert from apparent to absolute magnitudes we use the
cluster luminosity distance, correct the magnitudes for the Galactic
extinction (obtained from the maps of Schlegel et al. 1998), and apply
the K-correction of Fukugita et al. (1995) for elliptical galaxies,
which are likely to constitute the main cluster galaxy population.

The composite LF is obtained following Colless (1989) prescriptions. A
detailed description of the method can be found in paper II.

\begin{figure}
\begin{center}
\begin{minipage}{0.5\textwidth}
\resizebox{\hsize}{!}{\includegraphics{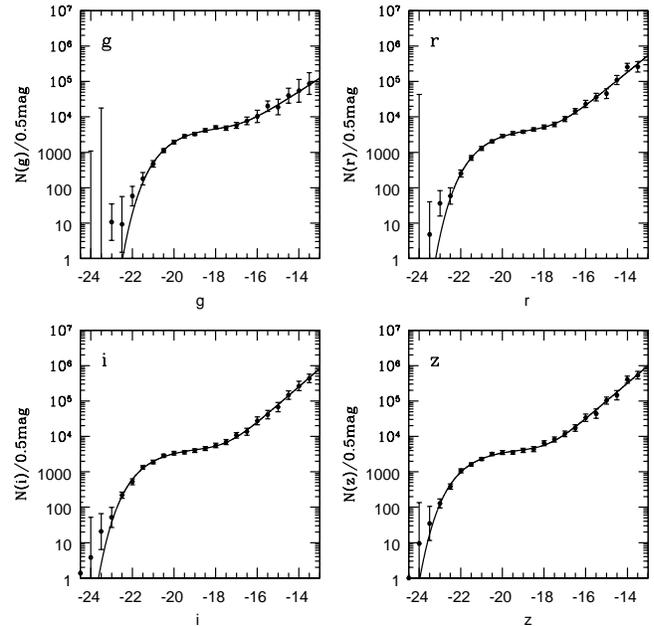}}
\end{minipage}
\end{center}
\caption{The 4 panels show the composite LFs in the 4 Sloan
bands. The individual LFs used to calculate the composite LFs are
measured within the physical sizes of the clusters, as given by
$r_{200}$.}
\label{funzione_di_lum}
\end{figure}

\subsection{Low surface brightness selection effect}
It is well known that magnitude-limited surveys may be biased against
low-surface brightness galaxies (e.g. Phillips \& Driver 1995).  An
assessment of this bias for the SDSS-EDR and SDSS-DR1 has been done by
Cross et al. (2004), who compared these catalogs with the Millennium
Galaxy Catalog (Liske, Lemon, Driver et al. 2003), a deep survey
limited in surface brightness to 26 mag~arcsec$^{-2}$. Cross et
al. (2004) concluded that the incompleteness of SDSS-EDR is less than
5\% in the range of effective surface-brightness $21 \le \mu_e \le 25$
mag~arcsec$^{-2}$, and it is around 10\% in the range $25 \le \mu_e
\le 26$ mag~arcsec$^{-2}$.  In this paper, galaxies contributing to
the faint-end of the cluster LFs have magnitudes $18 \le r \le 21$. In
this magnitude range, 65\% of the objects have $\mu_e \le 23$ mag
arcsec$^{-2}$, 30\% have $23 <
\mu_e \le 24$ mag~arcsec$^{-2}$, and 5\% have $mu_e \ge 25$
mag~arcsec$^{-2}$.  Hence, from the results of Cross et al. (2004), we
do not expect that the bias against low surface-brightness galaxies
results in an incompleteness above $\sim 5$\%. The faint-end of the
cluster LFs derived in this paper should thus be quite unaffected by
this selection effect. 

\begin{figure}
\begin{center}
\begin{minipage}{0.5\textwidth}
\resizebox{\hsize}{!}{\includegraphics{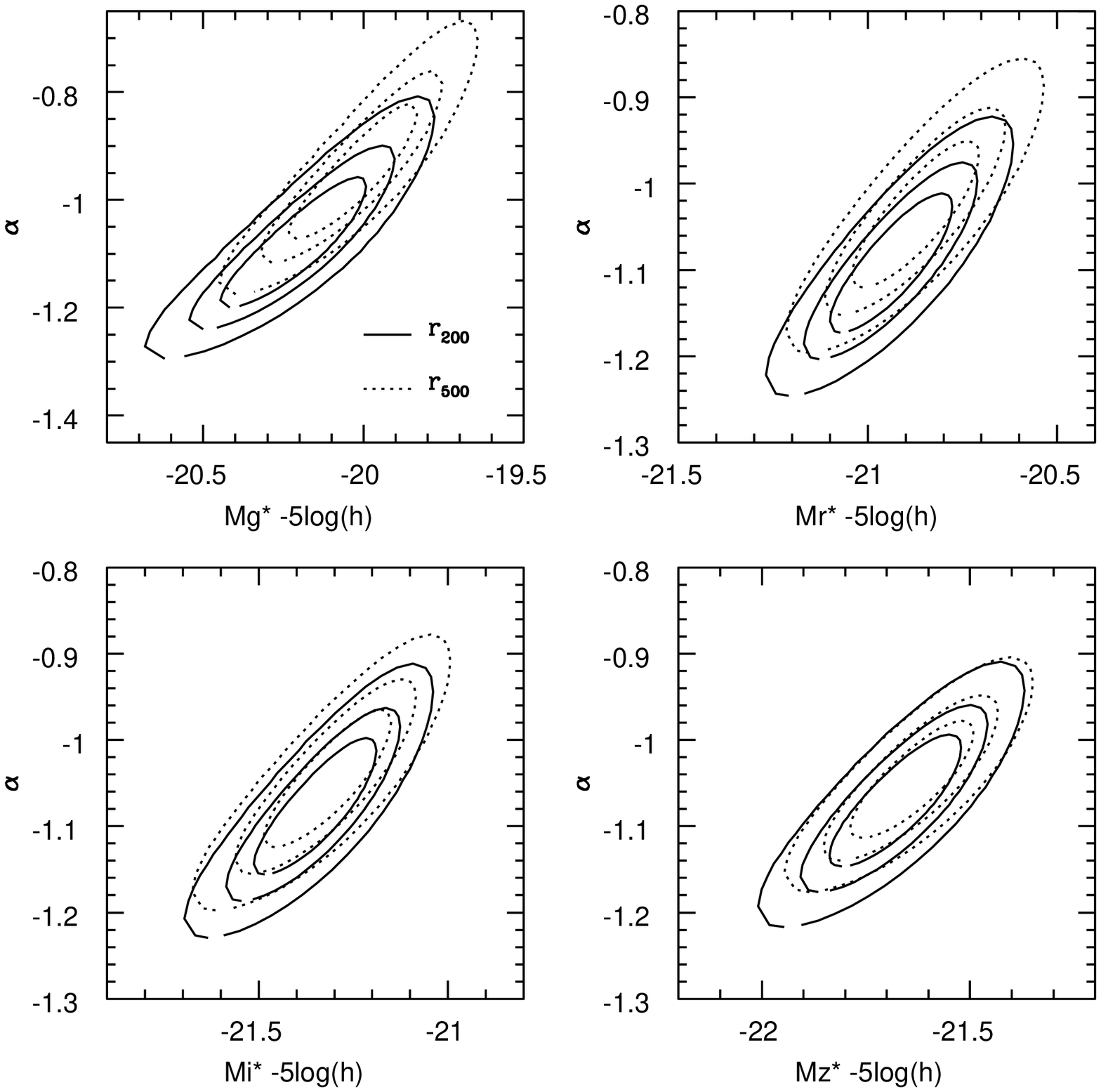}}
\end{minipage}
\end{center}
\caption{Contour plots of the 68\%, 95\%, and 99\% confidence levels
of the parameters of the bright-end component of the 
double-Schechter function fit to
the 4 SDSS bands composite LFs. Solid (dotted) contours show the
results for the composite LF computed within $r_{200}$ (respectively
$r_{500}$).}
\label{con1}
\end{figure}

\begin{figure}
\begin{center}
\begin{minipage}{0.5\textwidth}
\resizebox{\hsize}{!}{\includegraphics{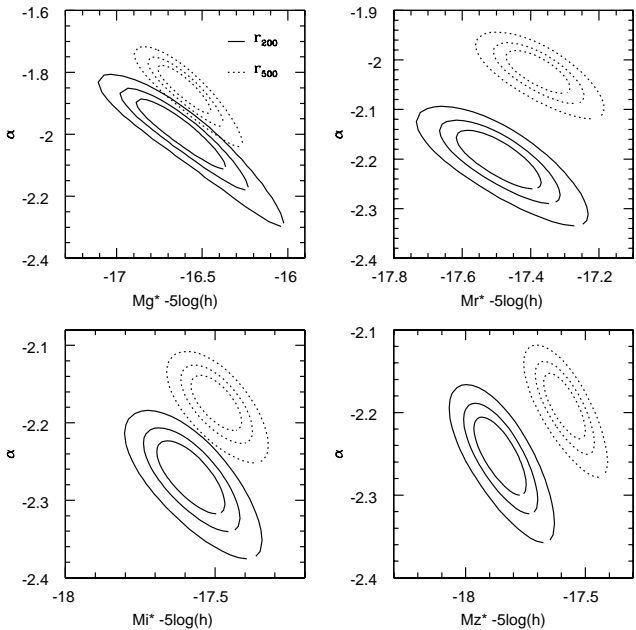}}
\end{minipage}
\end{center}
\caption{
Same as Fig.~\ref{con1}, but for the faint-end component.}
\label{con2}
\end{figure}

\section{The characteristic radii of galaxy clusters}
\label{s-glob}
We here describe the methods by which we measure the characteristic
radii $r_{500}$ and $r_{200}$. $r_{200}$ is the radius where the
mass density of the system is 200 times the critical density of the
Universe and it is considered as a robust measure of the virial radius
of the cluster. Similarly, $r_{500}$ is defined setting 500 instead
of 200 in the previous definition and it samples the central region of
the cluster.  Full details can be found in paper III.

We estimate a cluster characteristic radius through the virial
analysis applied on the redshifts of its member galaxies. We use
the redshifts provided in the SDSS spectroscopic catalog to define the
galaxy membership of each considered system. The SDSS spectroscopic
sample comprises all the objects observed in the Sloan r band with
pretrosian magnitude $r_P \le 17.77 $ mag and half-light surface
brightness $\mu_{50} \le 24.5$ mag$\rm{arcsec}^{-2}$. The SDSS DR2
spectrocsopic sample used for this analysis counts more tha 250.000
galaxies.

Cluster members are selected following the method of Girardi et
al. (1993). First, among the galaxies contained in a circle of radius
equal to the Abell radius, those with redshift $\mid cz - cz_{cluster}
\mid > 4000$ km s$^{-1}$ are removed, where $z_{cluster}$ is the mean
cluster redshift. Then, the gapper procedure (see also Beers et
al. 1990) is used to define the cluster limits in velocity
space. Galaxies outside these limits are removed. Finally, on the
remaining galaxies we apply the interloper-removal method of Katgert
et al. (2004; see Appendix A in that paper for more details).

The virial analysis (see, e.g., Girardi et al. 1998) is then performed
on the clusters with at least 10 member galaxies. The velocity
dispersion is computed using the biweight estimator (Beers et
al. 1990). The virial masses are corrected for the surface-pressure
term (The \& White 1986), using a Navarro et al. (1996, 1997) mass
density profile, with concentration parameter $c=4$. This profile
provides a good fit to the observationally determined average mass
profile of rich clusters (see Katgert et al. 2004).

Our clusters span a wide range in mass; since clusters of
different masses have different concentrations (see, e.g. Dolag et al.
2004) we should in principle compute the cluster masses, $M$'s, using
a different concentration parameter $c$ for each cluster.  According
to Dolag et al. (2004), $c \propto M^{-0.102}$. Taking $c=4$ for
clusters as massive as those analysed by Katgert et al. (2004), $M
\simeq 2 \times 10^{15} M_{\odot}$, Dolag et al.'s scaling implies our
clusters span a range $c \simeq 3$--6. Using $c=6$ instead of $c=4$
makes the mass estimates 4\% and 10\% higher at, respectively,
$r_{200}$ and $r_{500}$, while using $c=3$ makes the mass estimates
lower by the same factors. This effect being clearly much smaller than
the observational uncertainties, we assume the same $c=4$ in the
analysis for all clusters.

If $M_{vir}$ is the virial mass (corrected for the surface-pressure
term) contained in a volume of radius equal to the clustercentric
distance of the most distant cluster member in the sample, i.e. the
aperture radius $r_{ap}$, then,  the radius $r_{200}$ is then given by:
\begin{equation}
r_{200} \equiv r_{ap} \, [\rho_{vir}/(200 \rho_c)]^{1/2.4}
\label{e-r200}
\end{equation}
where $\rho_{vir} \equiv 3 M_{vir}/(4 \pi r_{ap}^3)$ and $\rho_c(z)$
is the critical density at redshift $z$ in the adopted cosmology. The
exponent in eq.(\ref{e-r200}) is the one that describes the average
cluster mass density profile near $r_{200}$, as estimated by Katgert
et al. (2004) for an ensemble of 59 rich clusters. Similarly,
$r_{500}$ is estimated by setting 500 instead of 200 in
eq.(\ref{e-r200}).  
%Finally, a $c=4$ Navarro et al. (1996) profile is used to
%interpolate (or, in a few cases, extrapolate) the virial mass
%$M_{vir,c}$ from $r_{ap}$ to $r_{200}$ and $r_{500}$.
 
\section{Analysis of the individual and composite LFs}
\label{s-lf}
In order to analyze the behavior of the composite LF faint-end as a
function of waveband and clustercentric distance, we define the number
ratio of dwarf to giant galaxies, DGR, as the ratio between the number
of faint ($-18 \le M \le -16.5$) and bright ($M<-20$) galaxies in the
cluster LF. The DGR is found to vary from cluster to cluster, more
than expected from statistical errors.  These variations are not
random however. As shown in Fig. \ref{sommario}, when the DGRs are
computed within a fixed metric radius, they are significantly
anti-correlated with several cluster global properties, i.e. the
cluster velocity dispersions, masses, and X-ray and optical
luminosities (velocity dispersions, virial masses, and X-ray
luminosities for our cluster sample were derived in paper III).  All
the correlations are very significant (1--$5 \times 10^{-5}$,
according to a Spearman correlation test). The more massive a cluster,
the lower its fraction of dwarf galaxies.

\begin{figure*}
\begin{minipage}{0.65\textwidth}
\resizebox{\hsize}{!}{\includegraphics{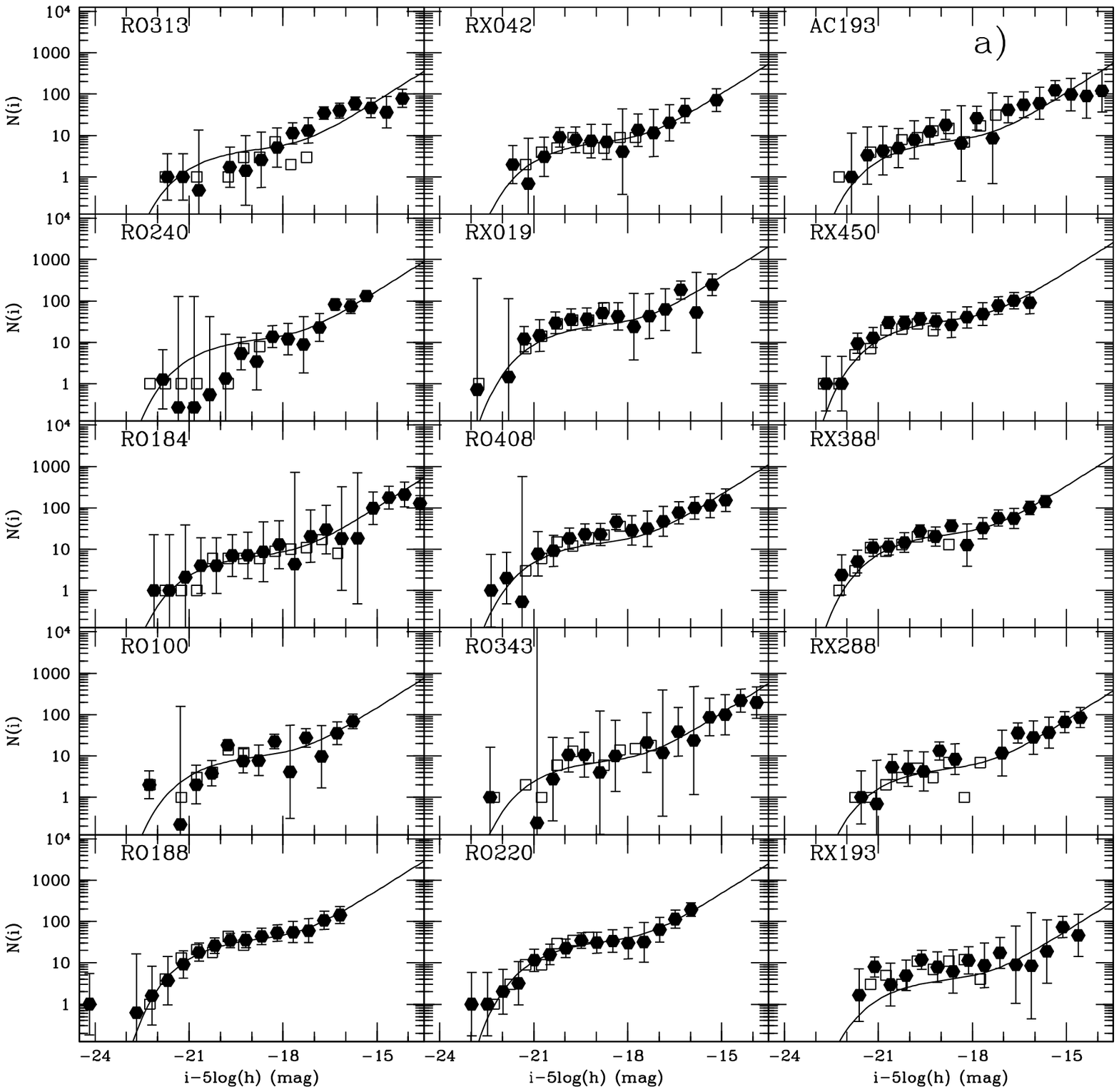}}
\end{minipage}
\begin{minipage}{0.28\textwidth}
\resizebox{\hsize}{!}{\includegraphics{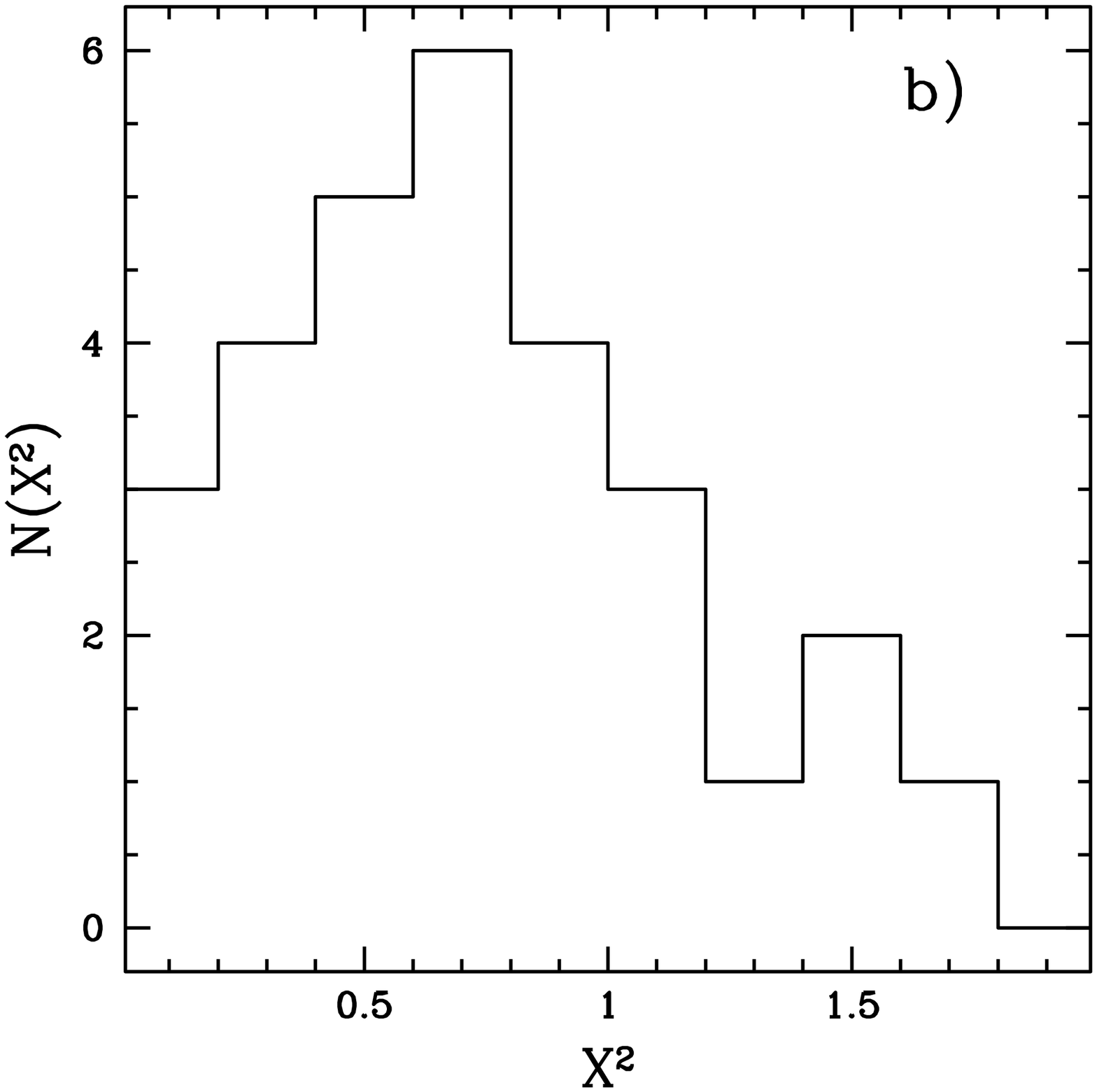}}
\resizebox{\hsize}{!}{\includegraphics{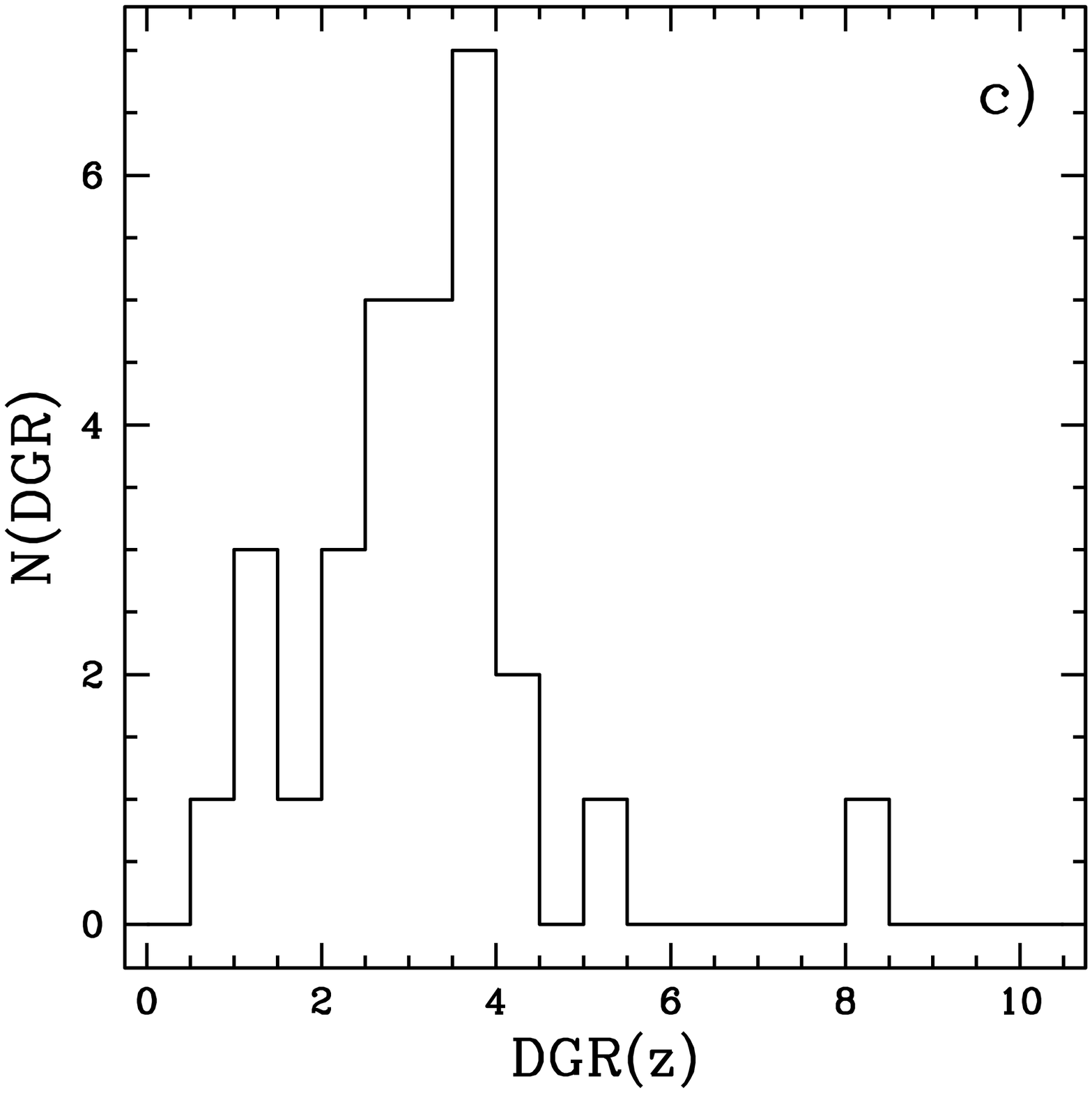}}
\end{minipage}
\caption{ Panel $a)$: the individual $r$-band LFs within $r_{200}$ of
a subsample of 15 clusters with the faintest absolute magnitude limit
($M_{r,lim}\le -15.5$).  Empty squares and filled points distinguish
the LFs computed from cluster members only (down to the SDSS
spectroscopic completeness magnitude, $r \le 17.77$), and using a
statistical background subtraction, respectively.  The solid line is
the composite LF. Cluster names are indicated. Panel $b)$: the
distribution of the $\chi^2$ values obtained from the comparison of
the composite and the 29 individual LFs of clusters with $M_{z,lim}
\ge -16.5$ mag.  Panel $c)$: the $z$-band DGR distribution of the 29
clusters.}
\label{15_cluster}
\end{figure*}

The correlation between cluster DGRs and cluster masses is most likely
due to the choice of a fixed metric aperture for all the clusters. In
fact, a fixed metric aperture samples larger (smaller) fractions of
the virialized regions of clusters of smaller (respectively, larger)
masses, and DGR is known to increase with clustercentric distance
(paper II).

\begin{table*}
\caption{Schechter parameters of the composite LF}
\begin{center}
\begin{tabular}[b]{c|ccccc}
\hline
  \multicolumn{1}{c}{}&\multicolumn{1}{c}{g}& \multicolumn{1}{c}{r}& \multicolumn{1}{c}{i} & \multicolumn{1}{c}{z} \\ \hline
\hline
\multicolumn{5}{c}{Double Schechter components function within $r_{200}$}\\ 
\hline
\hline
$\alpha_{b}$  &$-1.07 \pm 0.12$ & $-1.09\pm 0.09$ & $-1.08\pm 0.08$ & $-1.07 \pm 0.08$\\
$M^*_{b}$  &$-20.18\pm0.21$ & $-20.94\pm 0.16$ & $-21.35\pm 0.16$ & $-21.69\pm 0.15$\\
$\alpha_{f}$  &$-1.98 \pm 0.16$ & $-2.19\pm 0.09$ & $-2.26\pm 0.07$ & $-2.25 \pm 0.07$\\
$M^*_{f}$  &$-17.37\pm0.21$ & $-18.14\pm 0.15$ & $-18.43\pm 0.15$ & $-18.66\pm 0.14$ & \\
$\chi^2/\nu$ & 0.89 & 1.05 & 1.15 & 1.16 \\
\hline	
\hline
\multicolumn{5}{c}{Double Schechter components function within $r_{500}$}\\ 
\hline
\hline
$\alpha_{b}$  &$-0.97 \pm 0.09$ & $-1.05\pm 0.07$ & $-1.06\pm 0.06$ & $-1.05 \pm 0.05$\\
$M^*_{b}$  &$-20.04\pm0.15$ & $-20.84\pm 0.13$ & $-21.36\pm 0.14$ & $-21.67\pm 0.13$\\
$\alpha_{f}$  &$-1.84 \pm 0.11$ & $-2.02\pm 0.06$ & $-2.17\pm 0.05$ & $-2.19 \pm 0.06$\\
$M^*_{f}$  &$-16.61\pm0.22$ & $-17.38\pm 0.13$ & $-17.49\pm 0.12$ & $-17.58\pm 0.12$\\
$\chi^2/\nu$ & 0.87 & 0.98 & 1.11 & 1.09 \\
\hline
\hline
\multicolumn{5}{c}{Schechter+exponential  function within $r_{200}$}\\ 
\hline
\hline
$\alpha$  &$-0.88 \pm 0.25$ & $-1.26\pm 0.12$ & $-1.16\pm 0.13$ & $-1.16 \pm 0.12$\\
$M^*$  &$-19.95\pm0.29$ & $-21.16\pm 0.26$ & $-21.41\pm 0.22$ & $-21.71\pm 0.20$\\
$\beta$  &$-1.40 \pm 0.05$ & $-1.30\pm 0.07$ & $-1.26\pm 0.08$ & $-1.25 \pm 0.07$\\
$M^*_{t}$  &$-17.27\pm0.22$ & $-16.99\pm 0.43$ & $-17.65\pm 0.41$ & $-17.80\pm 0.39$\\
$\chi^2/\nu$ & 1.10 & 1.15 & 1.38 & 1.40 \\
\hline	
\hline
\multicolumn{5}{c}{Schechter+exponential  function within $r_{500}$}\\ 
\hline
\hline
$\alpha$  &$-0.88 \pm 0.25$ & $-1.05\pm 0.16$ & $-1.22\pm 0.14$ & $-1.00 \pm 0.14$\\
$M^*$  &$-19.94\pm0.29$ & $-20.91\pm 0.28$ & $-21.40\pm 0.25$ & $-21.54\pm 0.21$\\
$\beta$  &$-1.33 \pm 0.14$ & $-1.33\pm 0.09$ & $-1.22\pm 0.06$ & $-1.28 \pm 0.08$\\
$M^*_{t}$  &$-16.95\pm0.63$ & $-17.28\pm 0.50$ & $-17.43\pm 0.52$ & $-17.93\pm 0.45$\\
$\chi^2/\nu$ & 1.13 & 1.15 & 1.41 & 1.43 \\
\hline		
\end{tabular}
\end{center}
\label{tabella1}
\end{table*}

Because of this effect, the different cluster physical sizes must be
taken into account before comparing different cluster LFs. We then
determine the individual and composite LFs within $r_{500}$ and
$r_{200}$ for the subsample of 69 clusters of the RASS-SDSS galaxy
cluster sample for which these parameters are known (see paper III).

The composite LF calculated within $r_{200}$ is shown in
Fig.~\ref{funzione_di_lum} for four SDSS photometric bands.  The
$u$-band LF is not shown; in this band, there is no evidence for an
upturn at faint magnitude levels (see paper II). For all the other
bands LFs, a single Schechter function does not provide acceptable fits,
and a composite of two Schechter functions is needed:
\begin{equation}
\phi(L)=\phi^*[(\frac{L}{L^*_b})^{\alpha_b}exp(\frac{-L}{L^*_b})+(\frac{L}{L^*_f})^{\alpha_f}exp(\frac{-L}{L^*_f})]
\end{equation}
where $b$ and $f$ label the Schechter parameters of the
bright and faint end respectively.  From the reduced-$\chi^2$ values
given in Table \ref{tabella1} we conclude that a double-Schechter
function does provide adequate fits to the 4-bands composite LFs.
Alternatively, we fit the composite LFs with a function of this form:
\begin{equation}
\phi(L)=\phi^*(\frac{L}{L^*})^{\alpha}exp(\frac{-L}{L^*})[1+(\frac{L}{L_t})^{\beta}].
\end{equation}
In this function, $\phi^*$ , $L^*$ and $\alpha$ are the standard
Schechter parameters, $L_t$ is a transition luminosity between the two
power laws and $\beta$ is the power law slope of the very faint end
(Loveday 1997). Both functions require the same number of fit
parameters. However, the double Schechter component function provides
slightly better fits than the Schechter$+$power-law function in all
the Sloan bands (see Table \ref{tabella1}).

The Double Schechter function has been used for the first time
by Driver et al. (1994), while Thompson \& Gregory (1993) and Biviano
et al. (1995) suggested a Gaussian$+$Schechter function, to fit
respectively the bright and the faint end of the LF. More recently,
Hilker et al. (2003) used a double Schechter Function to fit the LF of
the Fornax cluster.

\begin{figure*}
\begin{center}
\begin{minipage}{0.48\textwidth}
\resizebox{\hsize}{!}{\includegraphics{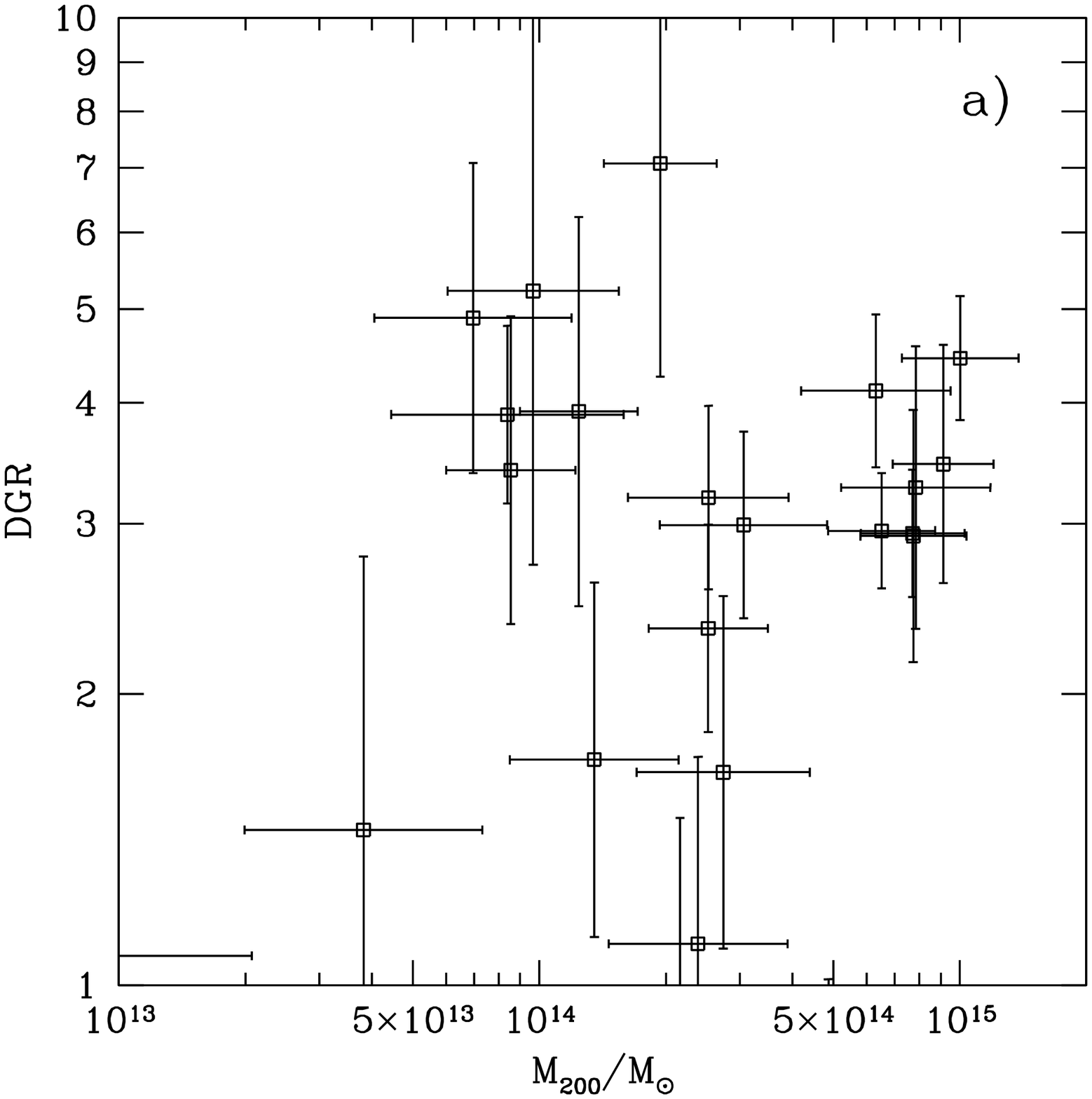}}
\end{minipage}
\begin{minipage}{0.48\textwidth}
\resizebox{\hsize}{!}{\includegraphics{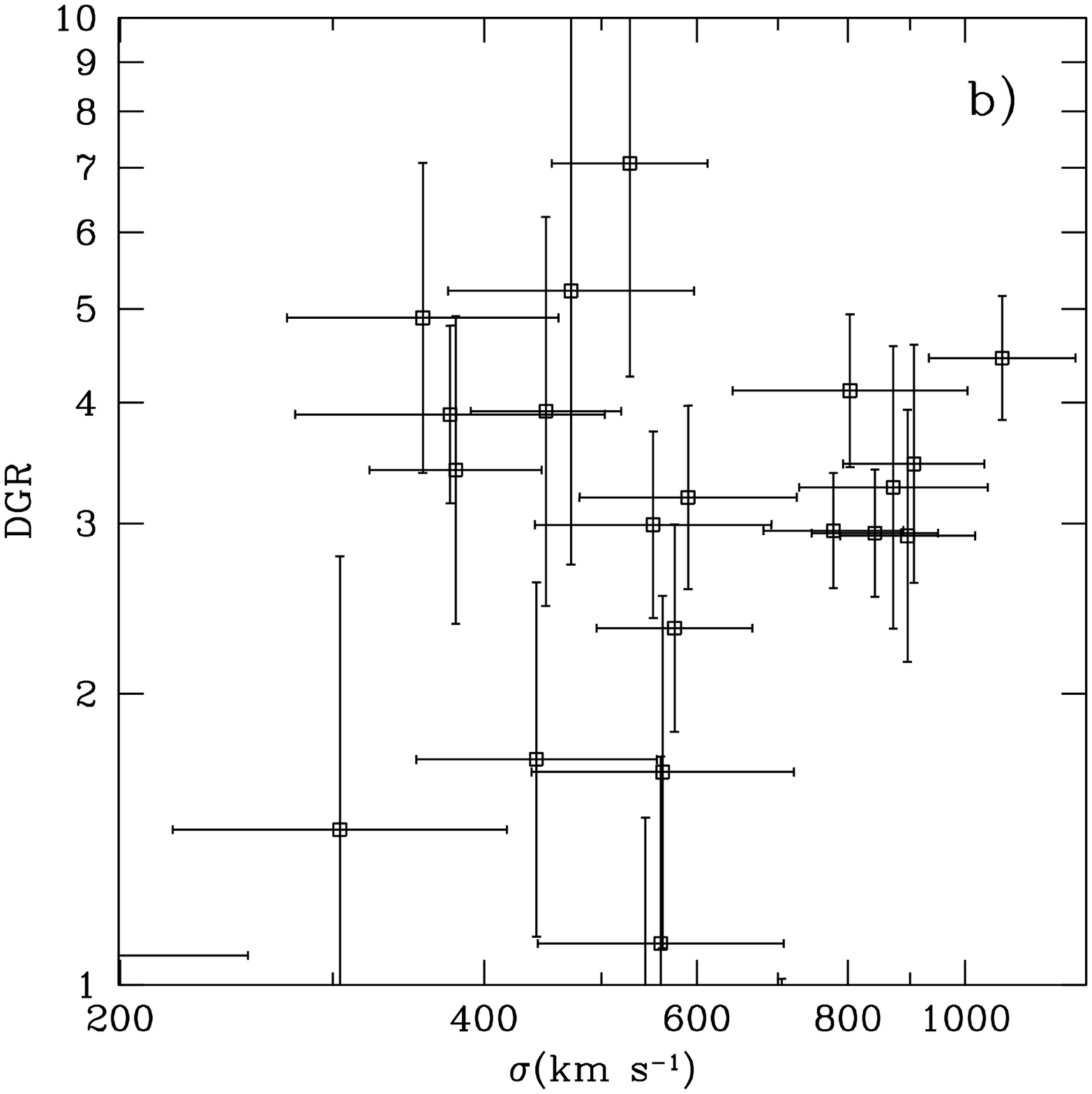}}
\end{minipage}
\end{center}
\caption{ 
The $z$-band DGR within $r_{200}$ as a function of cluster mass (panel
$a$) and the cluster velocity dispersion (panel $b$). If DGR is
calculated within $r_{200}$ the anti-correlation with mass ($\sigma$,
$L_X$ and $L_{op}$) disappears. }
\label{massona}
\end{figure*}

The confidence-level contours of the best-fit parameters of the
bright- and faint-end Schechter components are shown in
Figs. \ref{con1} and \ref{con2}, respectively. Both results for the
composite LF within $r_{500}$ (dotted contours) and $r_{200}$ (solid
contours) are shown. Clearly, the best-fit Schechter function to the
LF bright-end does not change significantly from $r_{500}$ to
$r_{200}$ (see Fig.~\ref{con1}) confirming the findings of paper II.
However, the faint-end LF steepens significantly (by 0.1--0.15 dex)
from $r_{500}$ to $r_{200}$, and the characteristic magnitude
correspondingly brightens by 0.3--0.4 magnitudes (see
Fig.~\ref{con2}), thereby indicating an increasing DGR with radius.
Our result is in agreement with the findings of paper II, and several
other works in the literature, which were however mostly based on
single cluster studies (e.g. Lobo et al. 1997; Durret et al. 2002;
Mercurio et al. 2003; Pracy et al. 2004; see however Trentham et
al. 2001, for a discordant result).

While our conclusions on the composite LF agree with those of paper
II, we find here different results concerning the individual cluster
LFs.  While in paper II we claimed significant LF variations from
cluster to cluster, we discover that such variations disappear when
the individual cluster LFs are computed within the physical sizes of
each cluster (defined by $r_{500}$ or $r_{200}$). This can be seen in
Fig.~\ref{15_cluster}a, where we plot the individual LFs of 15
clusters (those with the faintest absolute magnitude limits) and,
superposed, the composite LF, all measured within $r_{200}$ and in the
$r$-band.  The agreement between the composite and individual LFs is
very good.  Fitting the composite LF to the individual cluster LFs
result in the reduced-$\chi^2$ distribution shown in
Fig.~\ref{15_cluster}b.  For 90\% of the clusters the probability that
the composite and individual LFs are drawn from the same parent
distribution is larger than 95\%.

In Fig.~\ref{15_cluster}c we also show the $z$-band
DGR-distribution. When compared to the DGR distribution found in paper
II, the new DGR distribution is much narrower. In this paper we
considered the DGR within $r_{200}$ of 29 clusters, those with known
mass, $r_{200}$ and $r_{500}$, out of the 35 systems considered in
paper II. The mean value of the DGR is 3.5 and its dispersion is
indeed very close to the mean DGR statistical error of 1.4, as
expected if the individual cluster LFs are indeed all rather similar
when computed within a cluster-related physical radius.

Finally, in Fig.~\ref{massona} we show DGR within $r_{200}$ as a
function of the cluster mass $M_{200}$ (panel $a$) and the velocity
dispersion (panel $b$). There is no hint of the relation previously
found (compare with Fig.~\ref{sommario}a): the Spearman correlation
coefficient is $-0.08$, which is not statistically
significant. Similar results are found also for the $DGR-L_X$ and
$DGR-L_{op}$ relations.

Hence we conclude that the cluster to cluster LF variation seen in
paper II are entirely due to the use of a fixed metric aperture 
for all clusters, rather than an aperture sampling the same fraction
of the virialized region of each cluster.

\begin{figure*}
\begin{minipage}{0.5\textwidth}
\resizebox{\hsize}{!}{\includegraphics{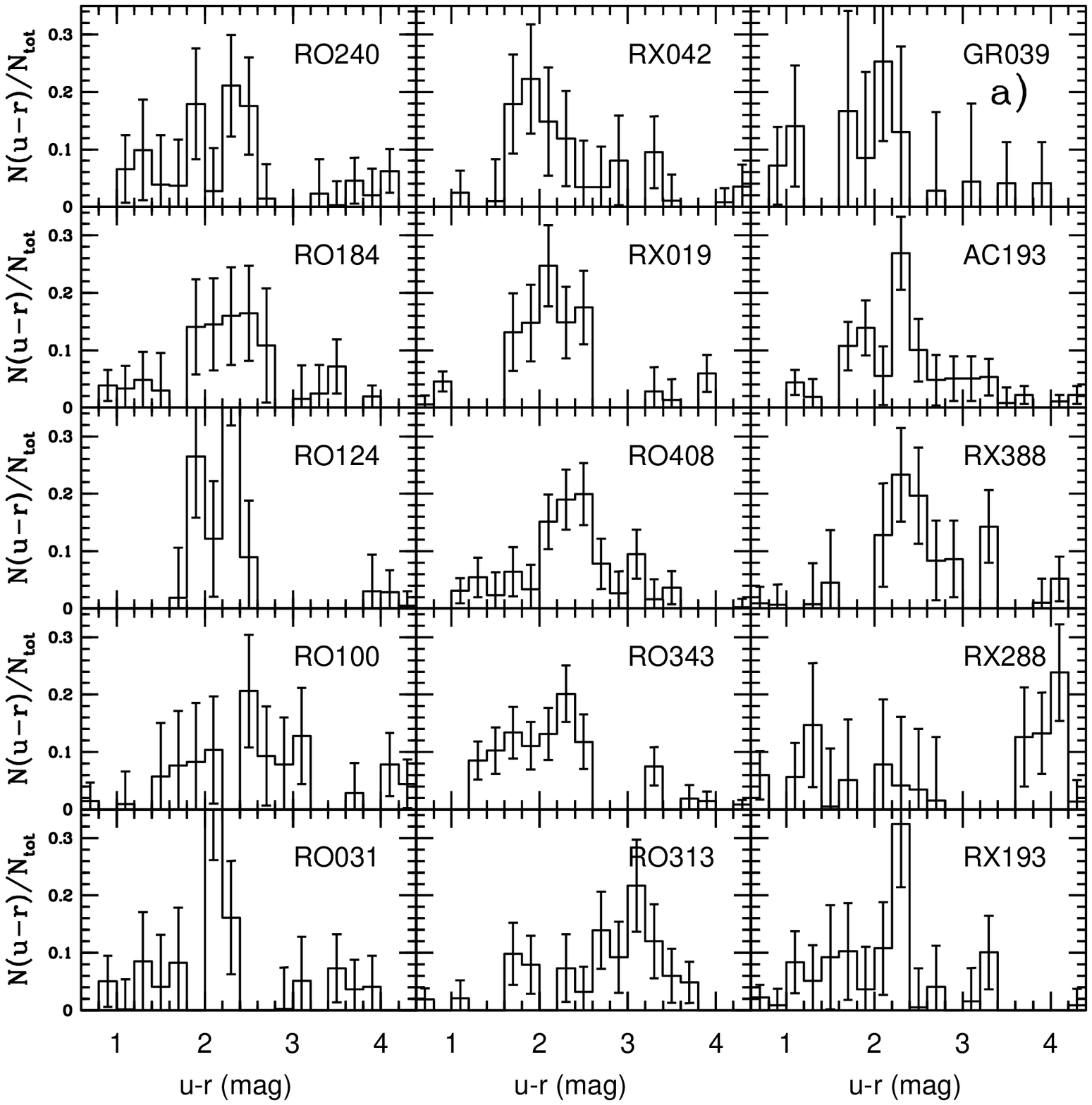}}
\end{minipage}
\begin{minipage}{0.5\textwidth}
\resizebox{\hsize}{!}{\includegraphics{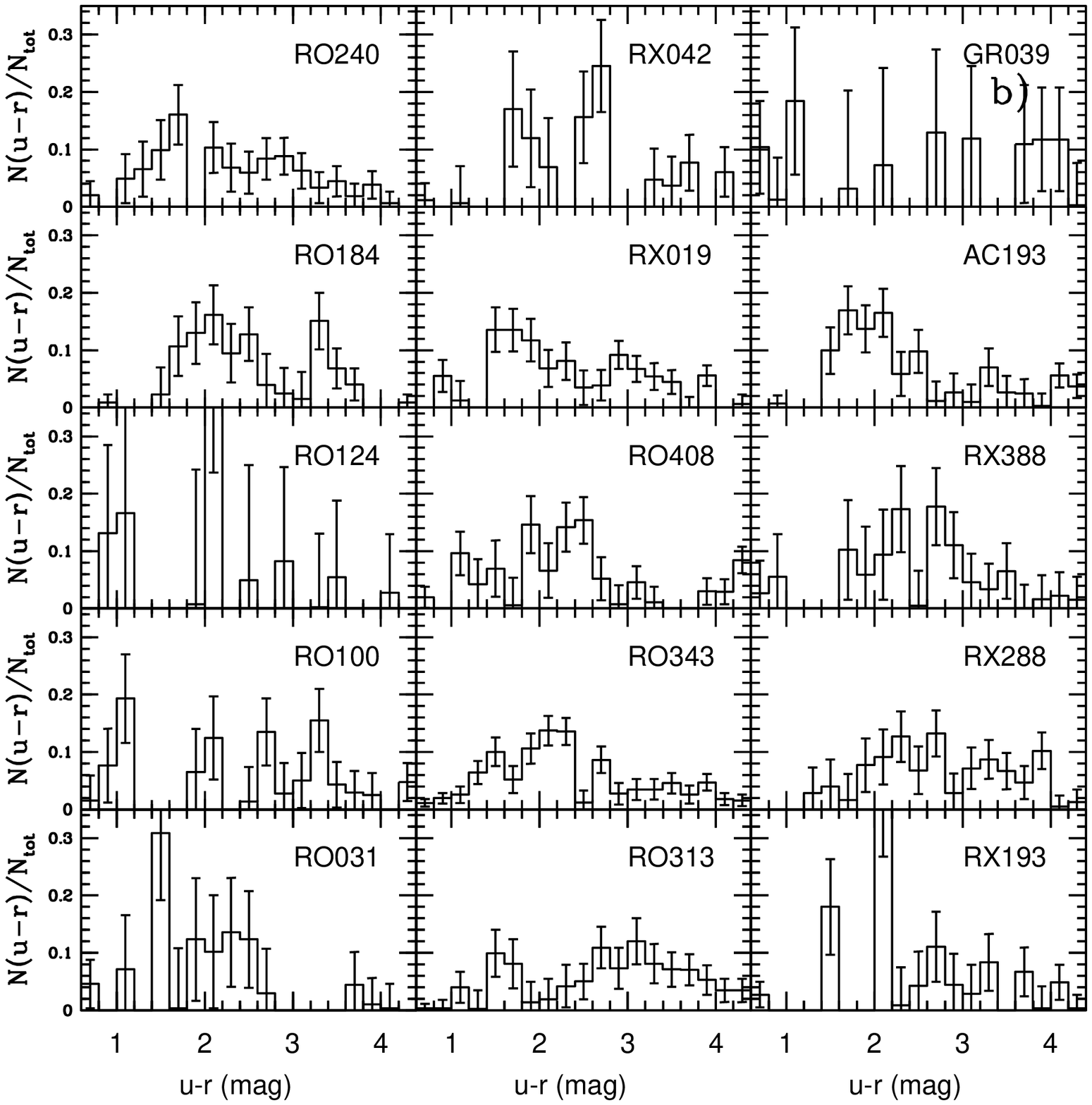}}
\end{minipage}
\caption{ The background-subtracted $u-r$ distribution of the galaxy
members of the 15 clusters with the faintest absolute magnitude limit
($M_{r,lim} \ge -15$). $a)$: color distribution in the magnitude range
$-18 \le M_r \le -16.5$; $b)$: color distribution in the magnitude
range $-16.5 \le M_r \le -15$.}
\label{coloru}
\end{figure*}

\section{The cluster LF per galaxy type}
\label{s-type}
In order to better understand the nature of the cluster galaxies
responsible for the LF upturn at low luminosities, we examine their
color distribution. In particular, we use the $u-r$ color, since the
$u-r$ distribution of Sa and earlier-type galaxies is well separated
from the $u-r$ distribution of Sb and later-type galaxies (Strateva et
al. 2001), thereby allowing to distinguish the two morphological
samples down to very faint magnitudes.

To define the color distribution of the cluster galaxies we
statistically subtract the contribution of field galaxies (Boyce et
al. 2001), using the same method applied for the statistical
subtraction of the background from the magnitude number counts.  We
determine the background color distribution of field galaxies in an
annulus around the cluster with inner radius larger than $r_{200}$;
significantly under- or over-dense regions (e.g. voids and background
clusters) are excluded. By subtracting the background color
distribution from the color distribution of galaxies in the cluster
region, we obtain the $u-r$ distribution of cluster galaxies.  The
validity of the method is confirmed by its application to the
spectroscopic subsample, for which cluster membership can be
established from the galaxy redshifts.

Fig.~\ref{coloru} shows the (background-subtracted) $u-r$
distribution of cluster galaxies in the range $-18 \le r \le -16.5$
(panel $a$) and $-16.5 \le r \le -15$ (panel $b$) for the subsample
of 15 clusters with the faintest absolute magnitude limit in the $r$
band ($M_{r,lim} \ge -15$). The error bars shown in the figure take
into account the galaxy counts Poisson statistics as well as the error
due to the background subtraction. 

At the redshifts of the 15 clusters considered ($0.02 \le z \le \
0.05$) early-type galaxies have $u-r$ colors in the range 2.6--2.9
(Fukugita et al. 1995), and galaxies redder than $u-r=3$ are probably
in the background.  Hence, we can see from Fig.~\ref{coloru}a that the
residual background contamination after the statistical background
subtraction, is generally small ($\le 10$~\%) and in fact not
significant in the bright magnitude range.  The contamination is
higher for the two clusters RO313 and RX 288, and probably due to the
presence of another cluster along the same line-of-sight.  In the
fainter magnitude range, the average background contamination
increases to 25--35\%, but is still not significant (see
Fig.~\ref{coloru}b).

\begin{figure}
\begin{center}
\begin{minipage}{0.5\textwidth}
\resizebox{\hsize}{!}{\includegraphics{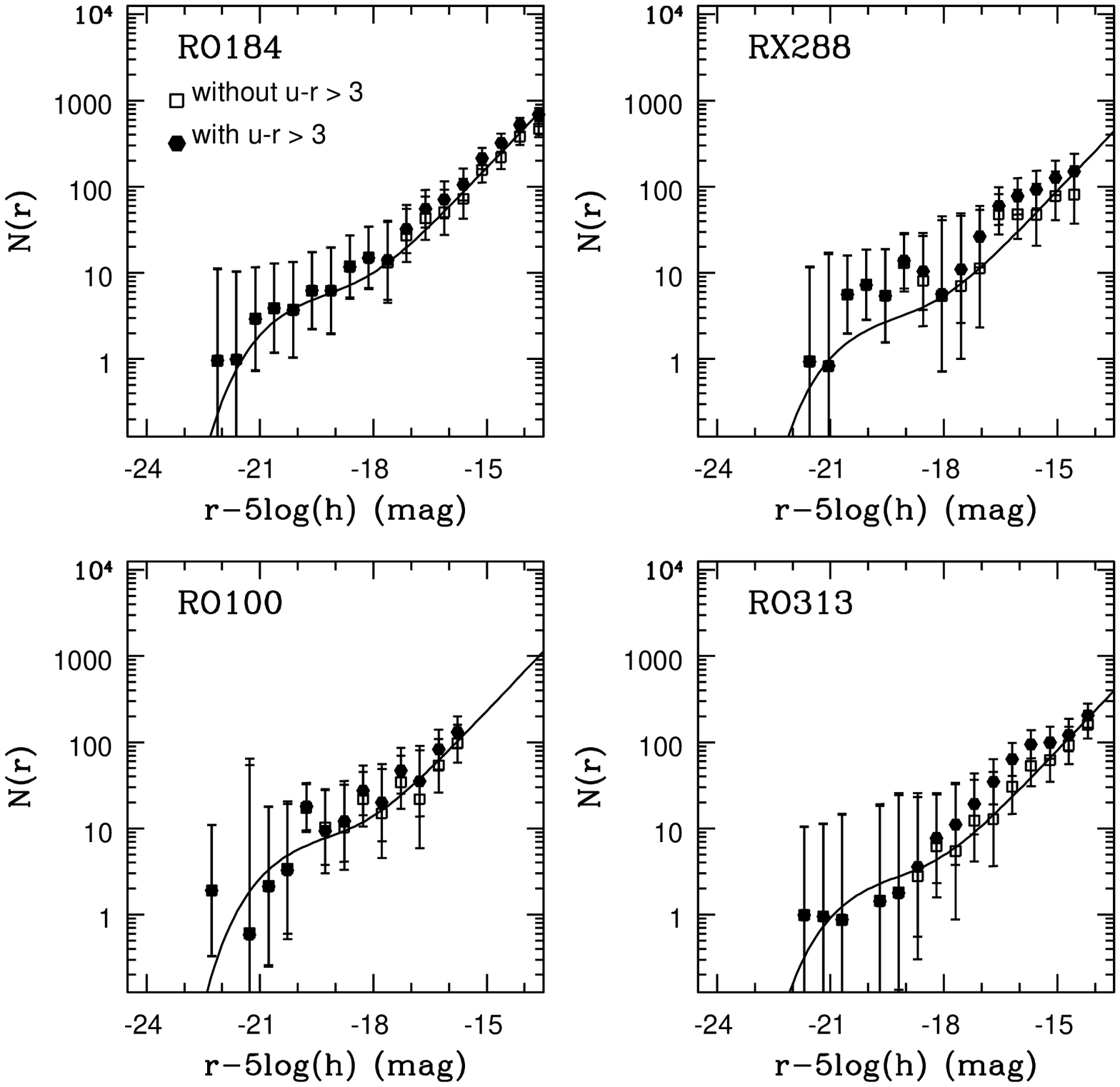}}
\end{minipage}
\end{center}
\caption{ The LFs of 4 clusters computed as for Fig.~\ref{15_cluster}
(filled points), and by additionally excluding all galaxies with $u-r
\ge 3$ (empty squares).}
\label{correction}
\end{figure}

If we exclude galaxies with $u-r \ge 3$ from our cluster samples, and
recalculate the cluster LFs as before (see \S~\ref{s-lfmethod}), the
modifications are marginal (compare filled points and empty squares in
Fig. \ref{correction}). If anything, a better agreement is now found
between the composite LF and the individual LF of the cluster R0313,
for which the background contamination is more severe, clearly
suggesting that the $u-r \ge 3$ color cut helps in cleaning the
cluster sample from background contamination.

\begin{figure*}
\begin{minipage}{0.5\textwidth}
\resizebox{\hsize}{!}{\includegraphics{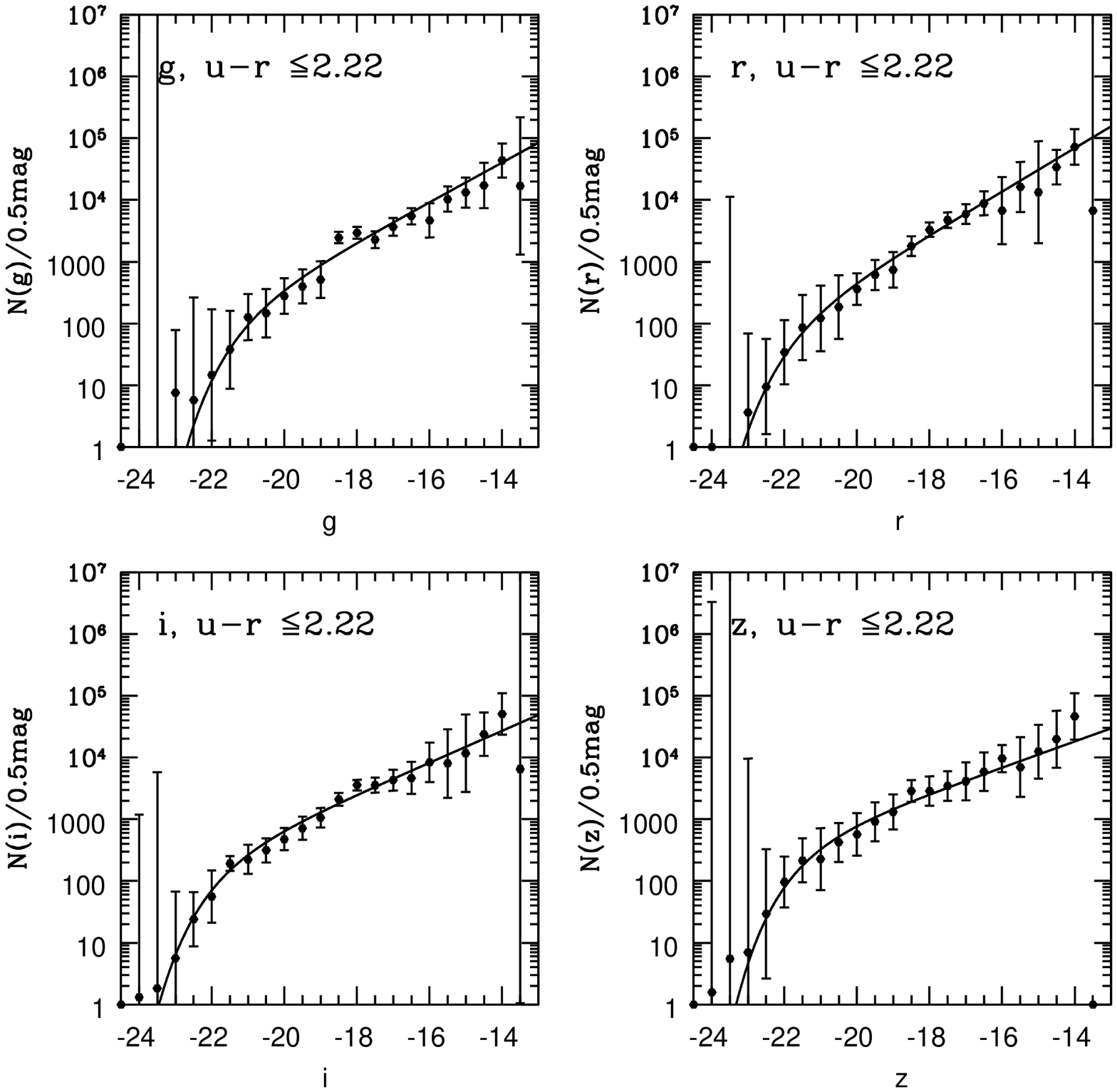}}
\end{minipage}
\begin{minipage}{0.5\textwidth}
\resizebox{\hsize}{!}{\includegraphics{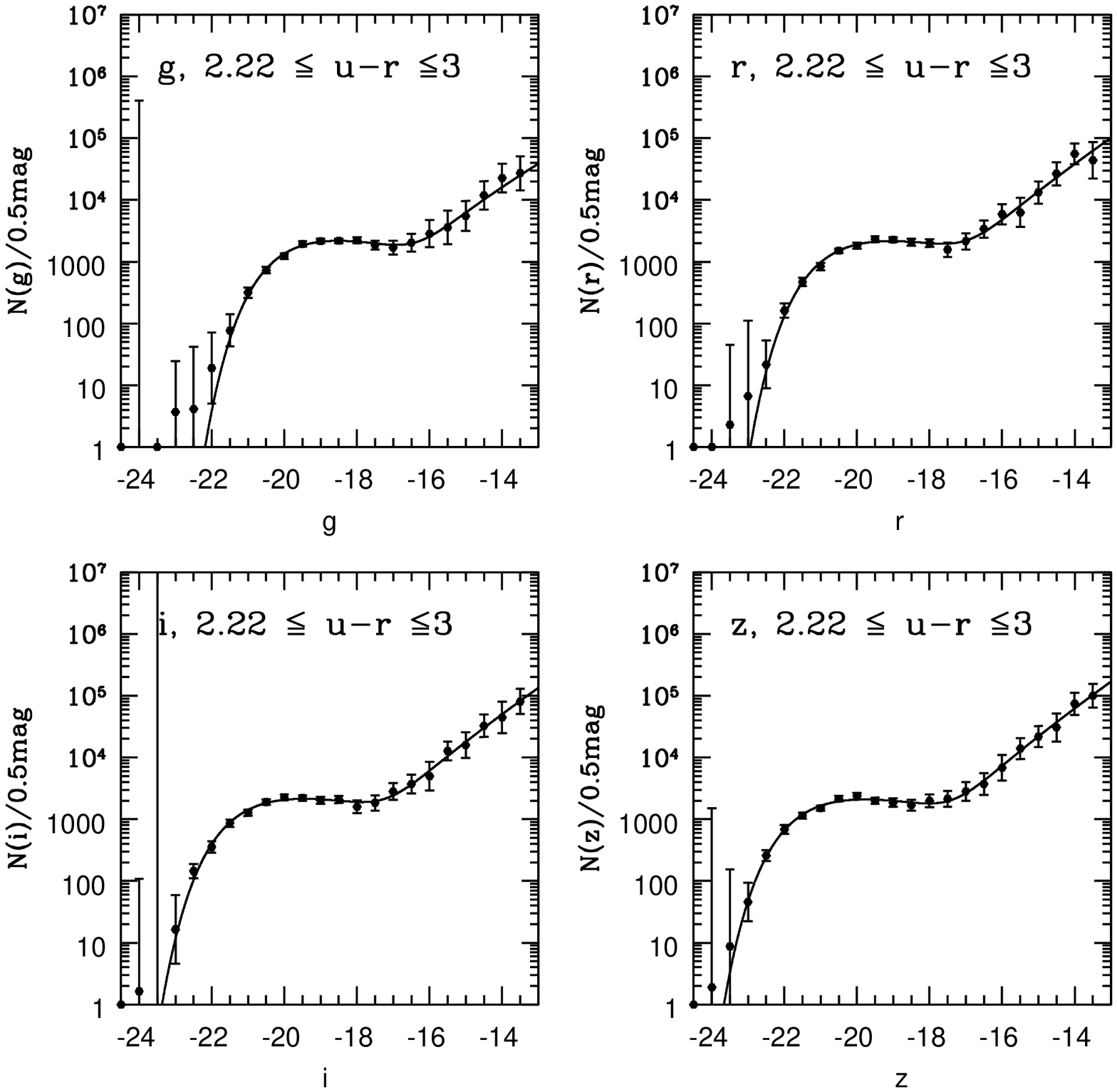}}
\end{minipage}
\caption{The composite late-type and early-type LFs in four SDSS
photometric bands.  The late-type (early-type) LFs are displayed in
the four panels on the left (respectively, right).}
\label{morphology}
\end{figure*}

We therefore adopt the $u-r < 3$ color cut to select cluster members,
and, following Strateva et al. (2001) we distinguish between cluster
early- and late-type galaxies using a color-cut $u-r=2.22$. We
restrict our analysis to the very nearby clusters ($z \le 0.1$) to
minimize the effects of an uncertain K-correction on the derived
colors. The composite LFs of the early- and late-type galaxies
(defined on the basis of their $u-r$ colors) are shown in
Fig.~\ref{morphology} for four SDSS photometric bands. The late-type
galaxy LF is well fitted by a single Schechter function and does not
show any evidence of an upturn at the faint end. On the other hand,
the early-type LF looks quite different from the late-type LF. It
shows a marked bimodal behavior with a pronounced upturn in the faint
magnitude region. The best fit parameters are listed in Table
\ref{tabella2}. Such an upturn is then reflected in the complete
(early$+$late) LF, with the late-type dwarf galaxies contributing to
make the faint-end of the complete LF even steeper. This result is in
agreement with Yagi et al. (2002).  They determine the total LF of 10
clusters within 1 h$^{-1}$ Mpc radius circle. They find that the
early-type LF cannot be fitted by a single Schechter function in the
magnitude range from $-23$ to $-15$, because it flattens at $M_R=-18$
and then rises again.

\begin{table*}
\caption{Schechter parameters of the early and late type galaxies composite LFs}
\begin{center}
\begin{tabular}[b]{c|ccccc}
\hline
  \multicolumn{1}{c}{}&\multicolumn{1}{c}{g}& \multicolumn{1}{c}{r}& \multicolumn{1}{c}{i} & \multicolumn{1}{c}{z} \\ \hline
\hline
\multicolumn{5}{c}{Double Schechter components ($r_{200}$) for early type galaxies}\\ 
\hline
\hline
$\alpha_{b}$  &$-0.69 \pm 0.10$ & $-0.75\pm 0.09$ & $-0.76\pm 0.09$ & $-0.76 \pm 0.08$\\
$M^*_{b}$  &$-19.79\pm0.16$ & $-20.57\pm 0.14$ & $-21.03\pm 0.15$ & $-21.30\pm 0.14$\\
$\alpha_{f}$  &$-1.86 \pm 0.15$ & $-2.01\pm 0.11$ & $-2.03\pm 0.08$ & $-2.05 \pm 0.09$\\
$M^*_{f}$  &$-17.37\pm0.21$ & $-18.14\pm 0.15$ & $-18.43\pm 0.15$ & $-18.66\pm 0.14$ & \\
$\chi^2/\nu$ & 0.72 & 1.04 & 1.03 & 0.90 \\
\hline	
\hline
\multicolumn{5}{c}{Single Schechter ($r_{200}$) for late type galaxies}\\ 
\hline
\hline
$\alpha$  &$-1.80 \pm 0.04$ & $-1.87\pm 0.04$ & $-1.64\pm 0.02$ & $-1.52 \pm 0.05$\\
$M^*$  &$-21.13\pm0.40$ & $-21.71\pm 0.52$ & $-21.79\pm 0.35$ & $-21.52\pm 0.47$\\
$\chi^2/\nu$ & 0.87 & 0.75 & 0.90 & 0.98 \\
\hline
\hline				
\end{tabular}
\end{center}
\label{tabella2}
\end{table*}

\begin{figure*}
\begin{center}
\begin{minipage}{0.49\textwidth}
\resizebox{\hsize}{!}{\includegraphics{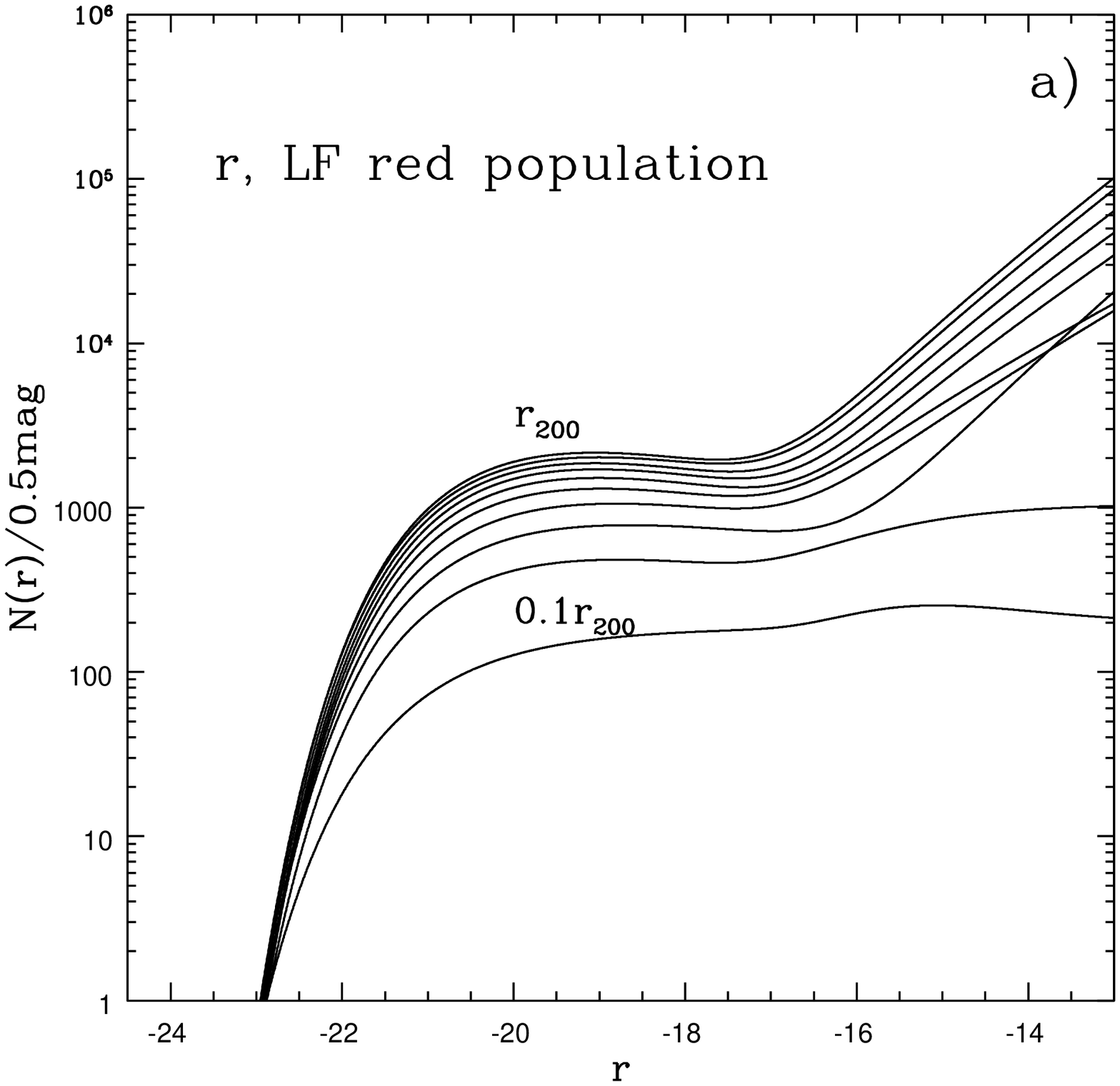}}
\end{minipage}
\begin{minipage}{0.49\textwidth}
\resizebox{\hsize}{!}{\includegraphics{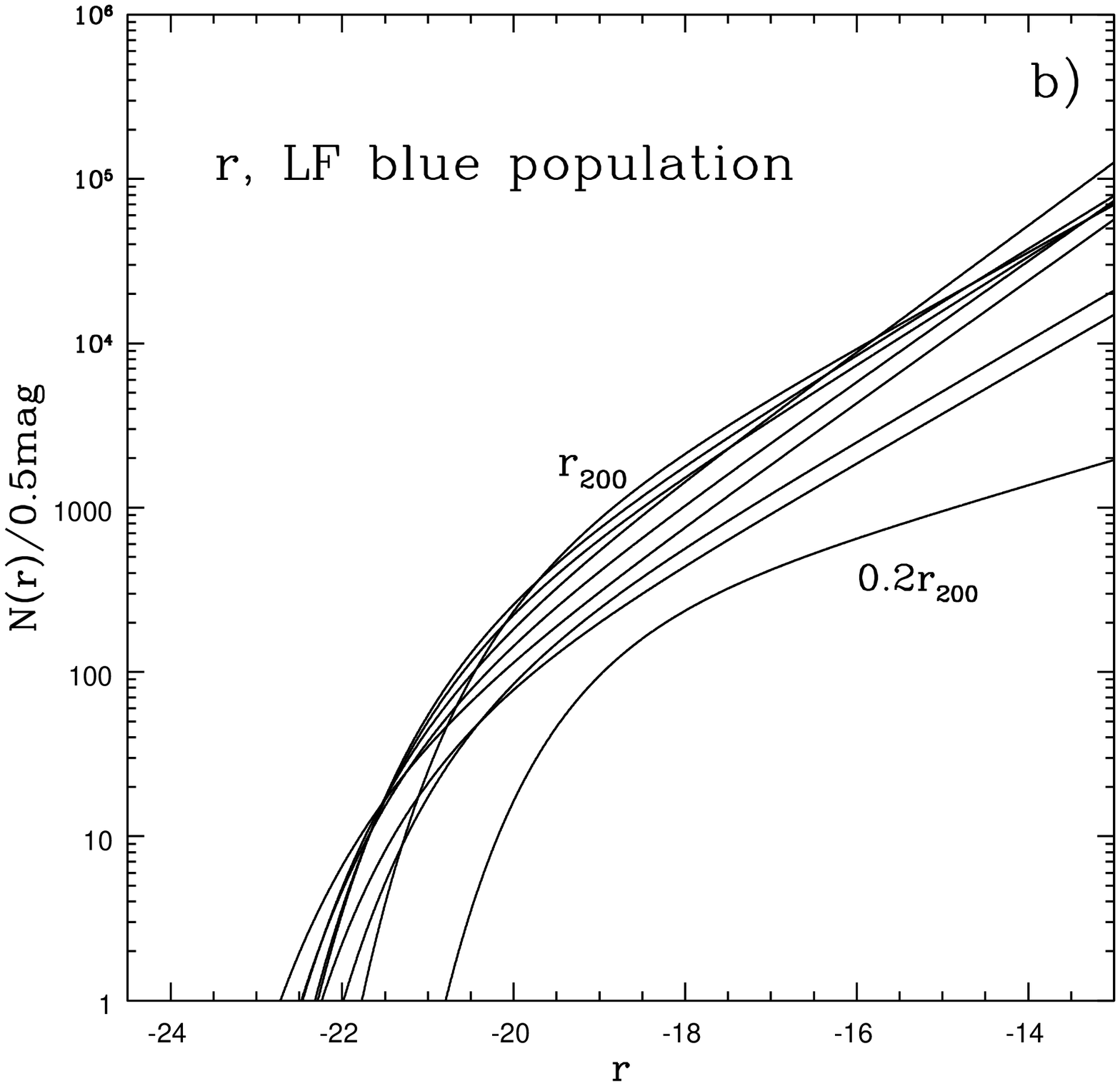}}
\end{minipage}
\end{center}
\caption{
The cluster LFs within different cluster apertures in the r band per
morphological type. The increment of the apertures is $0.1\times
r_{200}$. The normalization of the fitting function is increasing at
larger apertures. Panel $a)$ shows the LF of the cluster red galaxy
population, calculated within different clustercentric apertures
expressed in unit of $r_{200}$. Panel $b)$ shows the same for the
cluster blue galaxy population. For simplicity we only plot the best
fitting functions and not the data points.}
\label{alpha_spi}
\end{figure*}

\begin{figure}
\begin{center}
\begin{minipage}{0.5\textwidth}
\resizebox{\hsize}{!}{\includegraphics{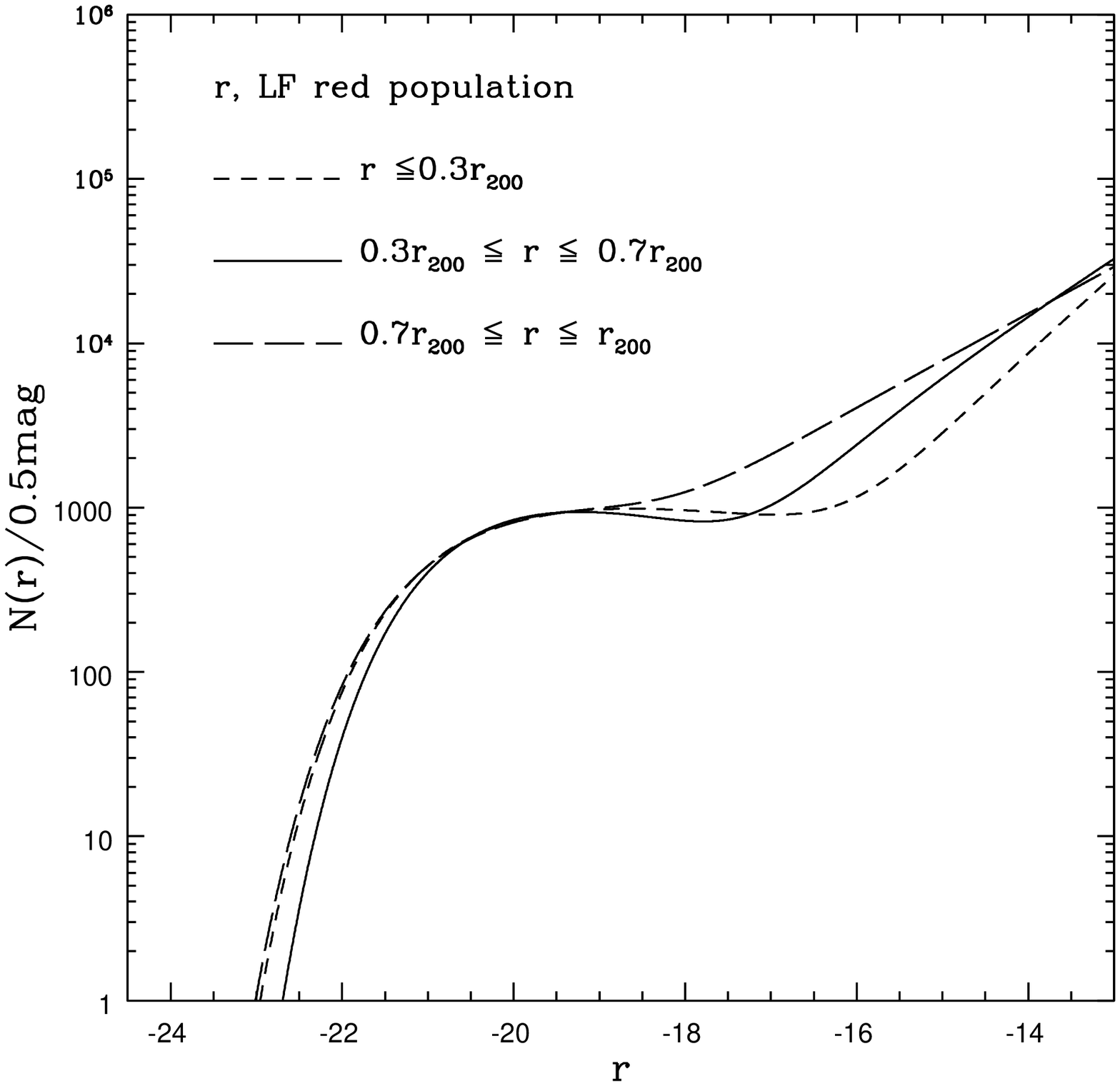}}
\end{minipage}
\end{center}
\caption{
The early-type LF calculated within three different cluster regions.
Only the best fitting functions are plotted, for simplicity,
and not the data points. The LFs are renormalized to the same value to
emphasize the shape variations.}
\label{gradiente}
\end{figure}

\section{The environmental dependence of the LFs}
\label{s-envi}
In order to gain insight into the processes responsible for the
shaping of the LF in clusters, we here examine the dependence of the
LF on the environmental conditions. In particular we analyze how the
LF shape, and the relative fraction of red and blue dwarf galaxies,
vary as a function of the clustercentric distance.
Fig. \ref{alpha_spi} shows the behavior of the cluster LF calculated
within different clustercentric apertures, separately for the
early-type (panel $a$) and late-type (panel $b$) galaxy populations.
Distances are in units of $r_{200}$.  For simplicity we only plot the
best fitting functions and not the data-points.  The early-type LF is
close to a Schechter function at the center of the cluster (within
$0.2 \; r_{200}$) and shows a marked upturn afterwards. The location
of the upturn varies from $-16.2 \pm 0.3$ mag at distances $\leq 0.3
\; r_{200}$ to $-17.4 \pm 0.4$ at distances $\leq r_{200}$. The
late-type LF is well fitted by a single Schechter function at any
clustercentric distance. We do not observe blue galaxies within $0.1
\; r_{200}$. Moreover, the central late-type LF at $0.2 \; r_{200}$ is
flatter than the LFs in the outer regions and shows a fainter
$M^*$. Since red galaxies are mostly high surface-brightness
objects (Blanton et al. 2004), the surface brightness selection effect
should be more important for the late-type LF, which, once corrected,
would become steeper at the faint-end. If anything, the difference in
slope between the faint-ends of the early- and late-type LFs should
thus be even larger than observed.

These results are confirmed by the analysis of the early-type LFs in
independent clustercentric rings.  We consider the region at
distances $r \le 0.3 \; r_{200}$ (the central ring), $0.3 \le
r/r_{200} \le 0.7$ (the intermediate ring) and $0.7 \le
r/r_{200} \le 1$ (the outer ring). The best fitting functions of
the cluster early-type LFs within these regions are shown in
Fig. \ref{gradiente}. In order to emphasize the {\em shape} variation
of the LF, all three LFs are renormalized to the same value.  The
upturn at the faint end is brighter in the outer ring than in the
central one, confirming the previous analysis. Moreover, the shape of
the bright end of the cluster LF seems to be absolutely independent
from the faint end. The values of $M^*$ and the slope of the bright
end are consistent within the errors in the three regions (as found in
paper II). This suggests that the process of formation of the bright
cluster galaxies (with magnitude brighter that $M^* -2$ mag) is the
same in all the cluster environments. Therefore, it seems unlikely
that the lack of dwarf systems observed at the center of the cluster
is due to a hierarchical process of formation of the bright central
galaxies. Indeed, in that case we should observe also a lack of bright
galaxies in the outer ring in favor of large amount of dwarf systems,
which is not observed.

\begin{figure*}
\begin{center}
\begin{minipage}{0.45\textwidth}
\resizebox{\hsize}{!}{\includegraphics{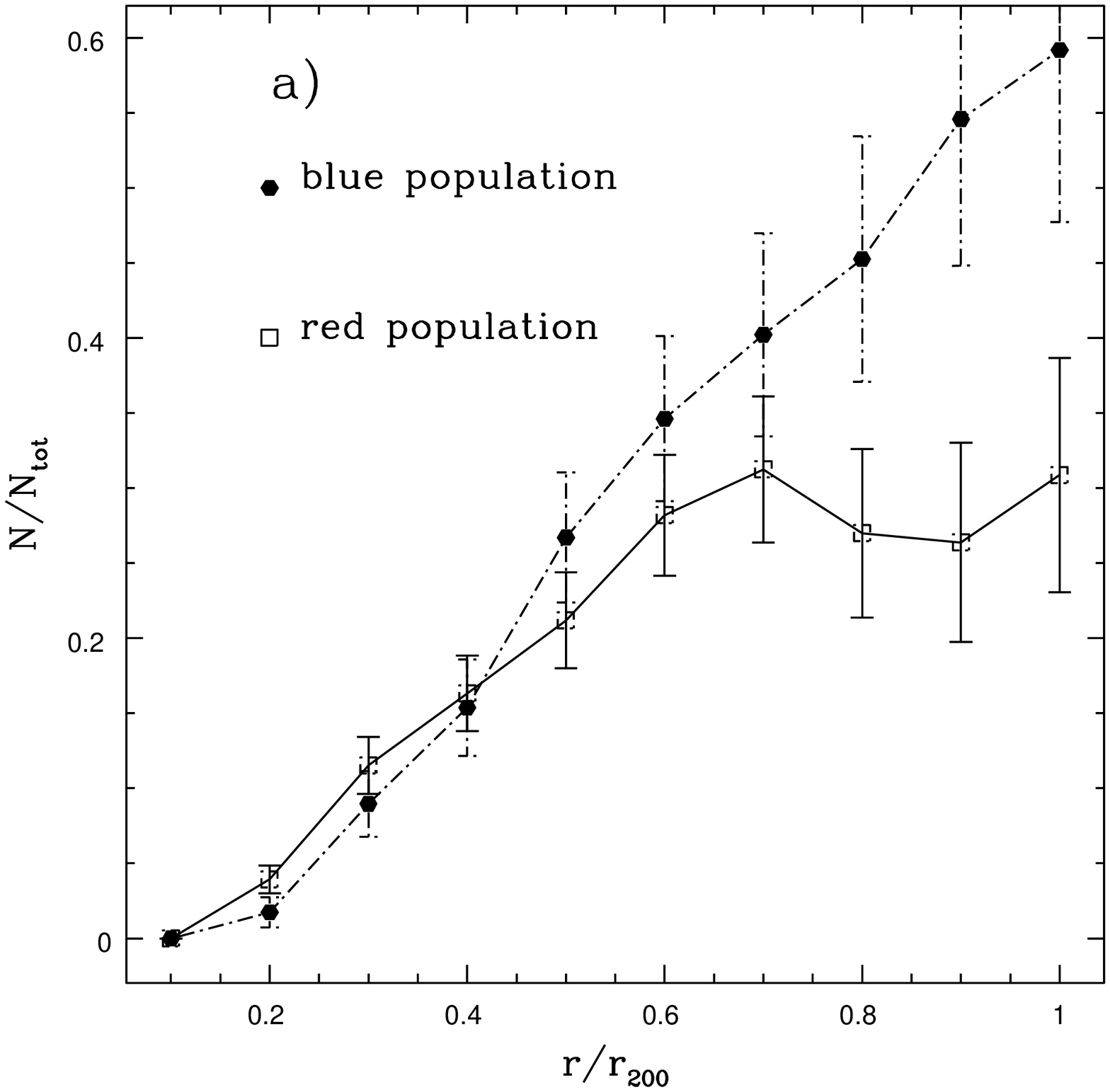}}
\end{minipage}
\begin{minipage}{0.45\textwidth}
\resizebox{\hsize}{!}{\includegraphics{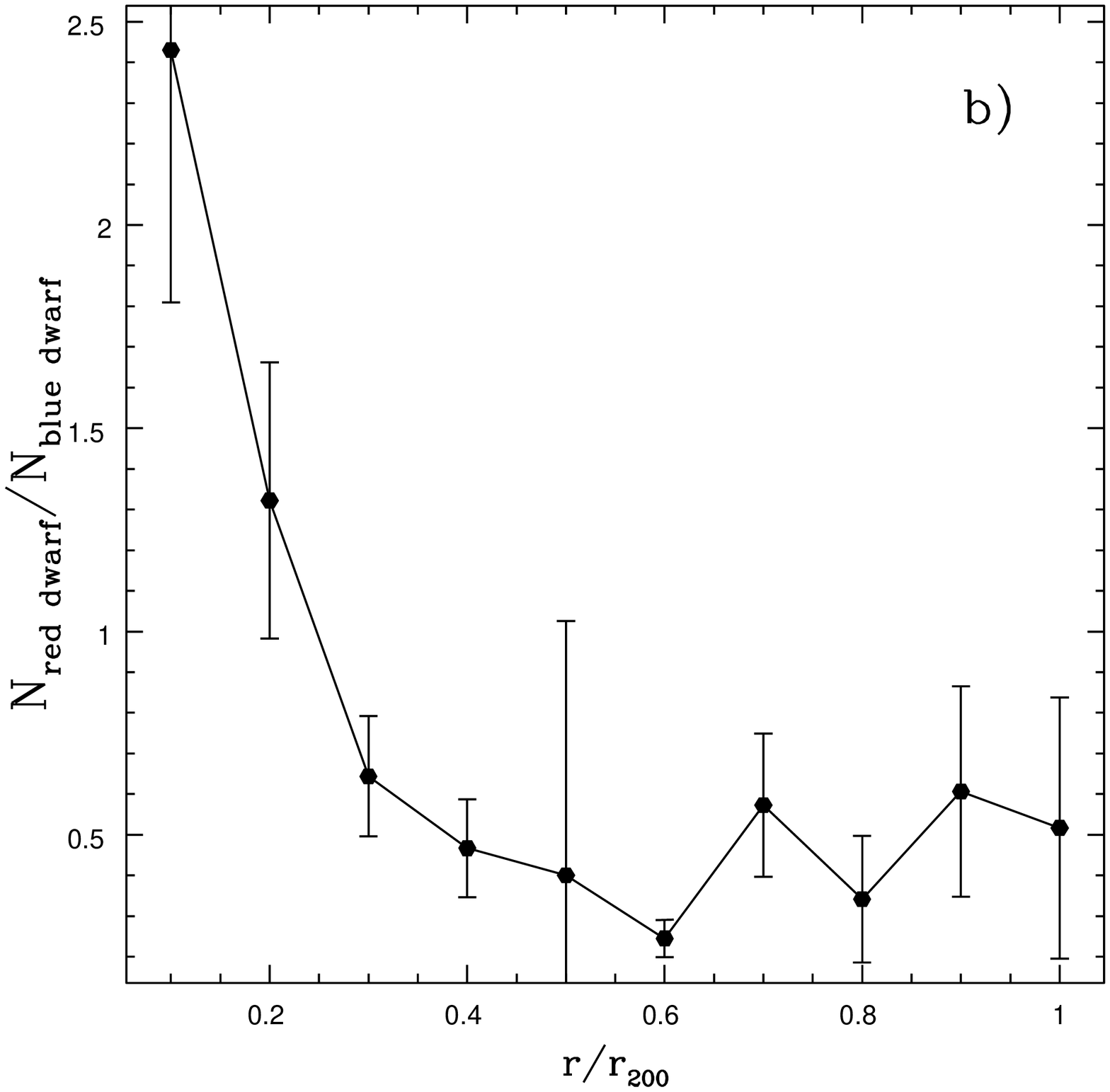}}
\end{minipage}
\begin{minipage}{0.45\textwidth}
\resizebox{\hsize}{!}{\includegraphics{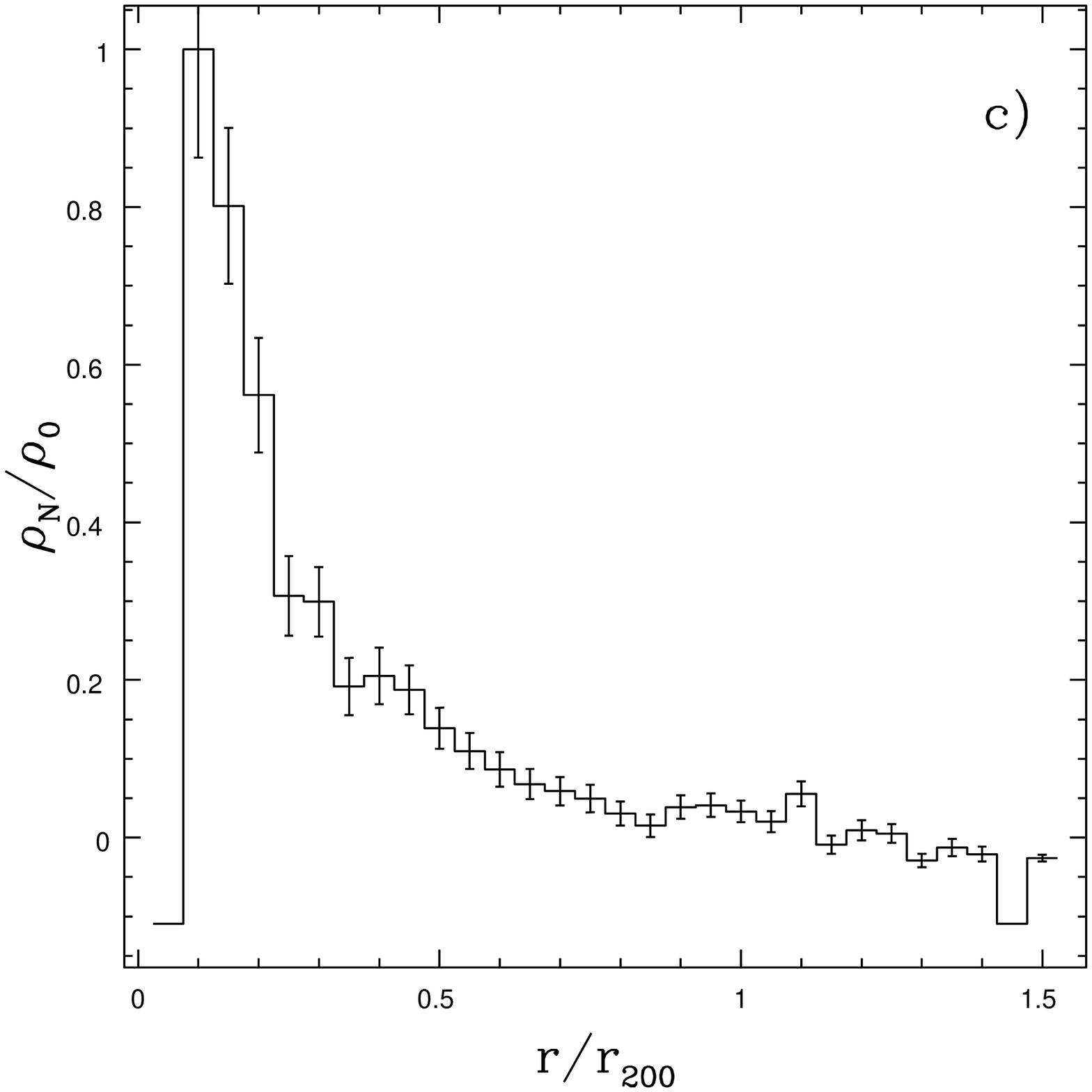}}
\end{minipage}
\begin{minipage}{0.45\textwidth}
\resizebox{\hsize}{!}{\includegraphics{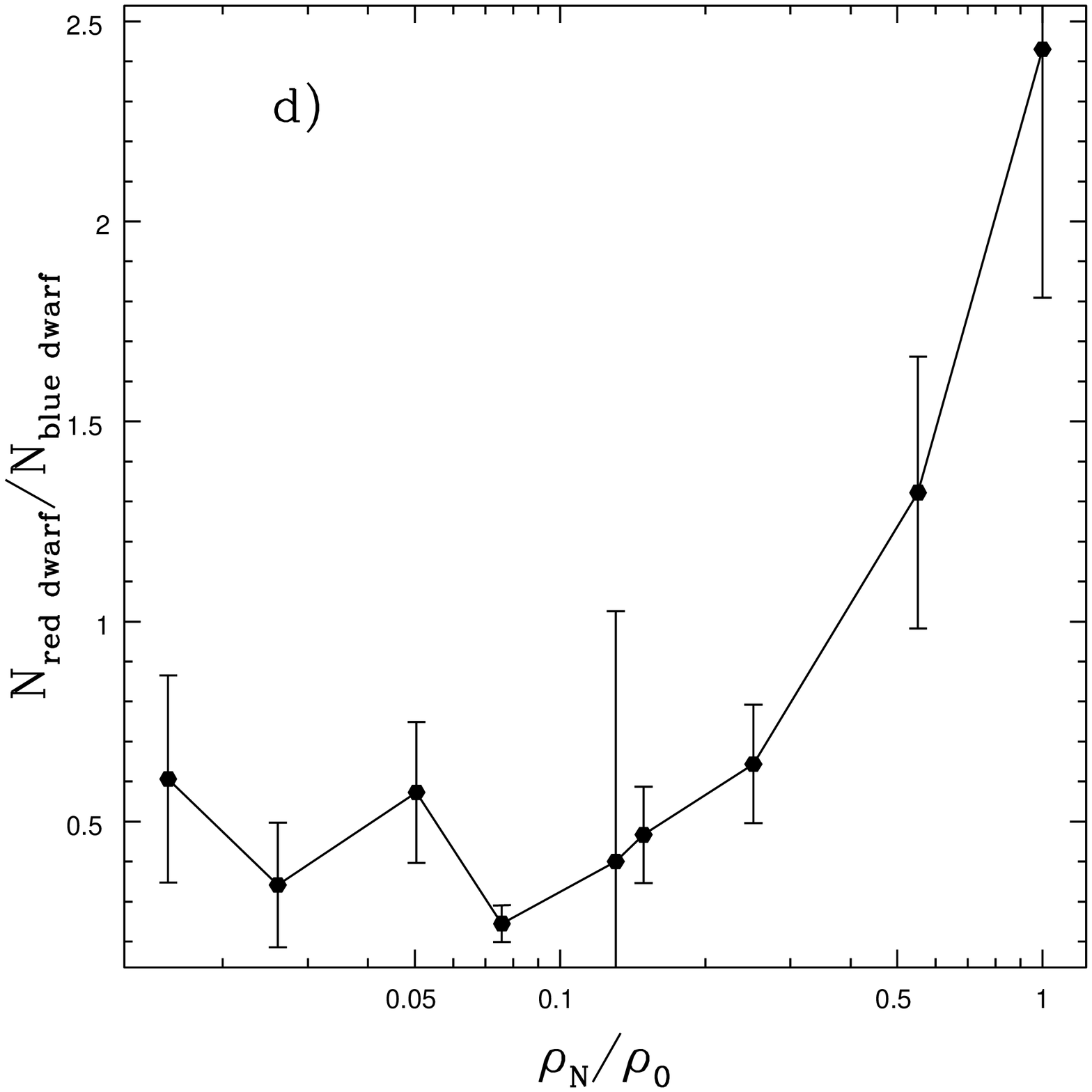}}
\end{minipage}
\end{center}
\caption{
The fraction of red and blue dwarf galaxies as a function of the
cluster environments. Panel $a)$ shows the cumulative radial profile
of the fraction of blue (filled points) and red (empty squared) dwarf
galaxies ($-18 \le M_r \le -15$ mag). The fraction is
defined on the total number of cluster dwarf galaxies in the
considered magnitude range. Panel $b)$ shows the differential radial
profile of the ratio between red and blue dwarf galaxies. Panel $c)$
shows the differential radial profile of the surface density of the
bright cluster galaxies in clusters ($M_r \le -18$ mag). Panel $d)$
shows the relation between the surface density of bright galaxies and
the fraction of red and blue dwarf galaxies calculated in the same
clustercentric ring.}
\label{profile}
\end{figure*}

The analysis so far provides only results about the LF {\em shape.} In
order to quantify the relative contribution of the early- and
late-type dwarf galaxy populations to the faint end of the LF, and its
dependence on the environment, we analyse the radial (cumulative and
differential) profile of the dwarf systems in the clusters. For this,
we consider the galaxies with $-18 \le M_r \le -15$, and to
improve the statistics, we stack the clusters with $M_{r,lim} \ge -15$
mag, by rescaling the clustercentric distances in units of $r_{200}$.
The cumulative profiles of the fractions of dwarf galaxies of both the
early- and the late-type are shown in Fig.~\ref{profile}$a$. The
center ($\leq 0.4 \; r_{200}$) contains less than 30\% of dwarf galaxies
(half of them are red systems), in the selected magnitude range. Dwarf
galaxies are more abundant in the cluster outskirts; the high-density
environment in the cluster cores is hostile to dwarf galaxies. This
phenomenology has already been observed in several individual clusters (see
e.g. Lobo et al. 1997; Boyce et al. 2001; Mercurio et al. 2003; Dahlen
et al. 2004);

The early-type dwarf galaxies
represents 35\% of the whole dwarf population within $r_{200}$, i.e.
most of the dwarf galaxies are of late-type. However, the dwarf
early-type galaxies are the dominant dwarf population region within
$0.4 \; r_{200}$, their relative fraction reaching a plateau at $\simeq
0.6 \; r_{200}$, while the late-type dwarf galaxies are more abundant in
clusters outskirts. This is confirmed also by the ratio between early-
and late-type dwarf galaxies calculated in contiguous clustercentric
rings (differential profile, see Fig.~\ref{profile}$b$). The number of
early-type dwarf galaxies is twice the number of late-type dwarf galaxies
within $0.2 \; r_{200}$ and then decreases to 1/2 at larger distances.

The relation between dwarf morphology and clustercentric distance
translates into a morphology-density relation.  In
Fig.~\ref{profile}$d$ we show the ratio between early- and late-type
dwarf galaxies as a function of the number density of galaxies
brighter than $M_r \le -18$ (the bright galaxies number density
profile is shown in panel $c$ of the same figure). As expected, the
early-type dwarf galaxies dominate in high density regions, while the
late-type dwarf galaxies are frequent in low density regions. Clearly,
the well known morphology density relation for cluster galaxies
(Dressler 1980) has an extension into the dwarf regime. 

\subsection{Comparison with the field}
\begin{figure}
\begin{center}
\begin{minipage}{0.5\textwidth}
\resizebox{\hsize}{!}{\includegraphics{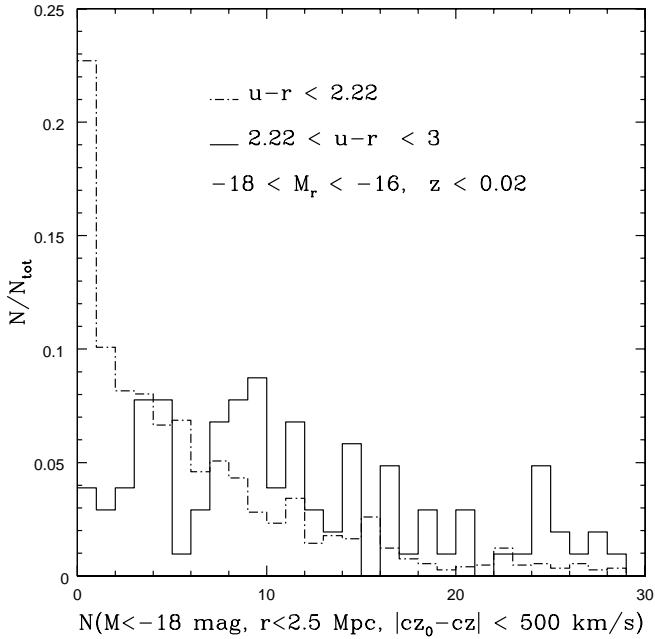}}
\end{minipage}
\end{center}
\caption{ 
The density distribution of neighbors of late
(dotted histogram) and early (solid histogram) galaxies in the
field. We select a fairly complete sample of nearby galaxies ($z\le
0.02$) from the Sloan spectroscopic sample in the magnitude region
$-18 \le M_R \le -16$. We calculate for each galaxy in the sample the
local density of neighbor galaxies counting the number of systems
with $M_R \le -18$ mag, within 2.5 Mpc projected radius and $\pm 500$
km/s of the galaxy position and redshift. The sample comprises 1561
systems.}
\label{N_dist}
\end{figure}

In order to extend the morphology-density relation for dwarf cluster
galaxies outside clusters, we extract a subsample of
galaxies from the SDSS spectroscopic sample. We select a fairly
complete sample of galaxies in the redshift range $z \le 0.02$ and in
the magnitude range $-18 \le M_r \le -16$. The late-type
galaxies ($u-r \le 2.22$) represent the 93\% of the galactic
population in that range of magnitude, in agreement with the results
of Blanton et al. (2004).  We then calculate for each galaxy in the
sample the local density of galaxy neighbors, by counting the number
of systems with $M_r \le -18$ mag, within 2.5 Mpc projected radius and
$\pm 500$ km/s of the galaxy position and redshift. We divide the
subsample in late and early-type galaxies using the color cut of
Strateva et al. (2001). Fig. \ref{N_dist} shows the number of galaxies
per bin of local density for the two galaxy types. It is clear that
late-type galaxies (dashed dotted histogram) populate the very low
density regions, while the early-type galaxies distribution (solid
histogram) has a much larger spread, with 50\% of the systems located
in regions with more than 10 galaxy neighbors. 

It is also interesting to compare our composite cluster LFs with the
LF of field galaxies.  Blanton et al. (2004) have recently derived the
LF of field SDSS galaxies down to $-12$ mag.  Their LF have a very
weak upturn, much shallower and at a fainter carachteristic magnitude
than in our cluster LF.  The faint-end slope of their LF is $-1.3$,
but could be steeper ($-1.5$) if a correction is applied
to account for low surface-brightness selection effects. The LF of
{\em blue} field galaxies is even steeper, but the authors do not report the
value of the faint-end slope. A similar faint-end slope ($-1.5$) has
also been found by Madgwick et al. (2002) for the LF of field galaxies
from the 2dF survey. They also noticed an upturn in the LF, due to an
overabundance of early-type galaxies, making it impossible to fit the
LF adequately with a single Schechter function. A previous
determination of the SDSS field LF was obtained by Nakamura et
al. (2003). They found a slope of $\sim -1.9$ for dIrr, consistent
with the value found by Marzke et al. (1994) for the CfA survey. 

The faint-end slope of our late-type cluster galaxies LF is steeper than
most field LFs for the same galaxy type (see Table~3 in Paper II) but
consistent with those of Nakamura et al. (2003) and Marzke et
al. (1994). Given the large variance of results for the field LFs,
possibly due to the different magnitude limits adopted, or to poor
statistics in the fainter bins of the LF (see de Lapparent 2003 for a
thorough discussion on this topic), we conclude there is no significant
difference between the late-type LF in clusters and the field.

\section{Discussion}
\label{s-disc}
There are many observations and theoretical models in the literature
that try to explain the formation and evolution of cluster galaxies,
red dwarf galaxies in particular. According to the hierarchical
picture for structure formation, small dark matter haloes form before
large ones. If one identifies the dwarf galaxies with the small dark
matter haloes, they are predicted to origin soon after the structure
formation began.  Dwarf ellipticals would then be old, passively
evolved galaxies.  This scenario seems to be inconsistent with the
observations of a large spread in age and metallicity in the clusters
dwarf early-type galaxies (Conselice et al. 2001,2003; Rakos et
al. 2001). Hence, dwarf ellipticals must have had a delayed star
formation epoch. The delay could be originated by the intense
ultraviolet background intensity at high redshift, keeping the gas of
the dwarf galaxies photoionized until $z\sim 1$, or, perhaps by the
intra-cluster medium confinement. The intra-cluster medium pressure
could avoid dwarf galaxies losing their gas content by SN
ejecta. However, this possibility would require a much more centrally
concentrated distribution of dwarf ellipticals in clusters than is
observed.

In alternative, the excess of dwarf early-type galaxies in clusters
could origin from the evolution of field dIrr when they are accreted
by the clusters.  The evolution of dIrr into dwarf early-type galaxies
is supported by the result of van Zee et al. (2004), namely that there
is significant similarity in the scaling relations and properties of
dIrr and dEs.  A scenario where {\em all} dwarf early-type galaxies
evolve from dIrr via disk fading does not however seem possible,
because many dEs in the Virgo and Fornax clusters are brighter than
the dIrr (Conselice et al. 2001).

Perhaps, some dwarf early-type galaxies evolve from dIrr and some
evolve from spirals. The evolution of spirals into dwarf spheroidals
can occur via the process of 'galaxy harassment' proposed by Moore et
al. (1996,1998). In this scenario, close, rapid encounters between
galaxies can lead to a radical transformation of a galaxy morphology.
Gas and stars are progressively stripped out of the disk systems,
eventually leaving a spheroidal remnant, that resembles an S0 galaxy
or a dwarf spheroidal, depending on the size of the progenitor. Direct
support for the harassment scenario comes from the discoveries of
disks or even spiral arms in dwarf early-type cluster galaxies (Jerjen
et al. 2000; Barazza et al. 2002; Graham et al. 2003). Indirect
support comes from the similar velocity distribution of dwarf cluster
galaxies (Drinkwater et al.  2001) and gas-rich spirals and irregulars
(Biviano et al. 1997), both suggesting infalling orbits.

Is the harassment scenario still viable in view of our results? We can
draw the following conclusions from our observational results. First,
the universality of the cluster LF suggests that whatever shapes the
cluster LF is not strictly dependent on the cluster
properties. Second, the difference between the cluster and field LF
seems to be related to an excess of dwarf early-type galaxies in
clusters. Hence, there is a cluster-related process that leads to the
formation of dwarf early-type galaxies, regardless of the cluster
intrinsic properties. The process cannot be related, e.g., to the intra-cluster
gas density, or the cluster velocity dispersion, or the cluster mass,
hence, a process like ram-pressure would seem to be ruled out.

The density dependence of the relative number of early- and late-type
dwarfs suggests that the shaping of the cluster LF is related to the
excess mean density relative to the field, which is the same for all
clusters if, as we have done, the cluster regions are defined within a
fixed overdensity radius ($r_{200}$ in our case). In other words, the
transformation of spirals, and perhaps, dIrr, into dwarf spheroidals
or dEs, seems to be a threshold process that occurs when the local
density exceeds a given threshold. Judging from Fig.~\ref{profile},
this threshold seems to occur at a clustercentric distance of $\sim
0.6$--0.7 $r_{200}$. 

We have also found that the relative number of dwarf early- and late-type
galaxies increases with decreasing clustercentric distance (and
increasing density). Galaxies near the cluster center are probably
an older cluster population, accreted when the cluster was smaller,
according to the hierarchical picture of cluster formation and evolution.
Hence, these centrally located galaxies have had more time to accomplish
the morphology transformation than galaxies located in the cluster
outskirts, which are more recent arrivals. 

On the other hand, very near the cluster center, an additional process
must be at work to explain our observed fading of the upturn of the
cluster early-type LF, and the decrease of both the early- and the
late-type dwarf-to-giant galaxy ratio with decreasing clustercentric
distance. High-velocity dispersions in clusters inhibit merging
processes (e.g. Mihos 2004), hence it is unlikely that dwarf galaxies
merge to produce bigger galaxies at the cluster centers. Consistently,
we find that the shape of the bright-end of the early-type LF does not
depend on the environment, which suggests that bright early-type
galaxies are not a recent product of the cluster environment. In fact,
the luminosity density profile of bright early-type galaxies has not
evolved significantly since redshift $z \sim 0.5$ (Ellingson 2003).

The most likely explanation for the lack of dwarf galaxies near the
cluster center is tidal or collisional disruption of the dwarf
galaxies. The fate of the disrupted dwarfs is probably to contribute
to the intra-cluster diffuse light (e.g. Feldmeier et al. 2004; Murante et
al. 2004; Willman et al. 2004).

The difference between the cluster and field LF could thus be
explained as a difference in morphological mix, plus a
density-dependent dwarf early-type galaxies LF, that, added to an
invariant bright early-type LF, produces a more or less important and
bright upturn, depending on the density of the environment.

\section{Conclusion}
\label{s-conc}
We have presented a detailed analysis of the cluster individual and
composite luminosity functions down to $-14$ mag in all the Sloan
photometric bands.  All the luminosity functions are calculated within
the physical size of the systems given by $r_{500}$ and $r_{200}$. The
main conclusions of our analysis are as follows:

\begin{itemize}
\item[-] We confirm that the composite LF shows a bimodal behavior
with a marked upturn at the faint magnitude range. A double Schechter
component function is the best fit for the cluster LF.  We show that
calculating the individual and the composite LF within a fixed
aperture for all the systems introduces selection effects.  These
selection effects justify the differences observed in the faint end of
the individual cluster LFs studied in paper II and the
anti-correlations between DGR and the global cluster properties (mass,
velocity dispersion, optical and X-ray luminosities) observed in this
work. If the cluster LF is calculated within the physical size of the
system ($r_{500}$ or $r_{200}$), the differences due to aperture
effects disappear and the individual cluster LF is well represented by
the composite LF.
Therefore, we conclude that the shape of the cluster LF
is universal in all the magnitude ranges.

\item[-]We use the $u-r$ color to study the color distribution of the
faint cluster galaxies. The color distribution confirms that the
contamination due to background galaxies is due to field-to-field
variance of the background. We apply the color cut at $u-r=2.22$
suggested by Strateva et al. (2001) to separate early-type from
late-type galaxies and study the composite LF by morphological
type. We observe that the upturn at the faint magnitudes shown by the
complete LF is due to early-type galaxies while the late-type LF is
well represented by a single Schechter function.

\item[-]We study the cumulative and the differential radial profile of
the faint early- and late-type galaxies in clusters. The faint early-type
galaxies are concentrated in the central regions while the faint late-type
galaxies dominate the outskirts of the systems.  The analysis of the
color-density relation in a reference sample of nearby galaxies
selected from the SDSS spectroscopic sample suggests that red
galaxies could be a typical cluster galaxy population. Our analysis
show that the bright red population seems to have a luminosity
distribution absolutely independent from the behavior of the faint red
galaxies in different environments. We observe a fading of the LF
upturn toward the cluster core. 

\item[-]We propose to interpret our results in term of a combination
of two processes, transformation of spirals and dIrr into dwarf
early-type galaxies via harassment, and disruption of dwarf galaxies
near the cluster center by collisions and/or tidal effects.
\end{itemize}

Whether galaxies evolve from one type to another, in response to the
local density, to create the morphology-density relation, or whether
the relation is established when the galaxies form, is still an open
issue (see, e.g., Dressler 2004). Photometric data alone cannot
provide conclusive indications about the nature and the origin of the
dwarf population in cluster. In this respect, it would be very useful
to sample the velocity distributions of a large set of dwarf galaxies
in clusters, in order to constrain their orbital characteristics as it
has recently been done for bright cluster galaxies (Biviano \& Katgert
2004). If the dwarf early-type galaxies evolve from spirals, radially
elongated orbits are expected, while if dwarf early-type galaxies are
a more pristine cluster population, their orbits should resemble the
isotropic orbits of ellipticals. Additional insights may come from
higher accuracy spectroscopy of the dwarf galaxies, allowing to deduce
information about their internal velocity dispersion and metallicity,
which could be used to put constraints on their age (see, e.g.,
Kauffmann et al. 2004; Carretero et al. 2004).

\vspace{1cm}

\begin{acknowledgements}
We would like to thank the anonymous referee for the usefull
comments which significantly improved the paper.
\end{acknowledgements}

Funding for the creation and distribution of the SDSS Archive has been
provided  by    the  Alfred P.   Sloan   Foundation, the Participating
Institutions, the  National Aeronautics and  Space Administration, the
National  Science  Foundation, the  U.S.    Department of  Energy, the
Japanese Monbukagakusho, and the Max Planck Society. The SDSS Web site
is   http://www.sdss.org/. The  SDSS is  managed  by the Astrophysical
Research Consortium   (ARC) for the   Participating Institutions.  The
Participating Institutions  are The  University of Chicago,  Fermilab,
the Institute for Advanced  Study, the Japan Participation  Group, The
Johns  Hopkins    University,  Los Alamos   National  Laboratory,  the
Max-Planck-Institute for   Astronomy (MPIA), the  Max-Planck-Institute
for   Astrophysics (MPA), New Mexico  State  University, University of
Pittsburgh, Princeton University, the United States Naval Observatory,
and the University of Washington.


\begin{thebibliography}{}
\bibitem[Abazajian et al.(2003)]{dr1}
Abazajian, K., Adelman-McCarthy, J., Ag\"ueros, M., et al. 2003, AJ, 126, 2081 
\bibitem[Andreon (2004)]{andre}
Andreon, S. 2004, A\&A, 416, 865
\bibitem[Barazza et al.(2002)]{bara}
Barazza, F.D., Binggeli, B., \& Jerjen, H. 2002, A\&A, 391, 823
\bibitem[Beers et al.(1990)]{beers2}
Beers, T.C., Flynn, K., \& Gebhardt, K. 1990, AJ, 100, 32
\bibitem[Beijersbergen et al.(2002)]{bei}
Beijersbergen, M., Hoekstra, H., van Dokkum, P.G., \& van der Hulst, T. 2002, MNRAS, 329, 385
\bibitem[Biviano et al. (1995)]{biv95}
Biviano, A., Durret, F., Gerbal, D., et al. 1995, A\&A, 297, 610
\bibitem[Biviano \& Girardi (2003)]{bivgir}
Biviano, A., \& Girardi, M. 2003, ApJ, 585, 205
\bibitem[Biviano \& Katgert (2004)]{bivkat}
Biviano, A., \& Katgert, P. 2004, A\&A, 424, 779
\bibitem[Biviano et al.(1997)]{biviano1}
Biviano, A., Katgert, P., Mazure, A., et al. 1997, A\&A, 321, 84
\bibitem[Blanton et al.(2003)]{blanton}
Blanton, M.R., Lin, H., Lupton, R.H., et al. 2003, AJ, 125, 2276
\bibitem[Blanton et al.(2004)]{blanton1}
Blanton, M.R., Lupton, R.H., Schlegel, D.J., et al. 2004, astro-ph/0410164, submitted to AJ
\bibitem[B\"ohringer et al.(2000)]{bh1}
B\"ohringer, H., Voges, W., Huchra, J.P., et al. 2000, ApJS, 129, 435
\bibitem[B\"ohringer et al.(2002)]{bh2}
B\"ohringer, H., Collins, C.A., Guzzo, L., et al.  2002, ApJ, 566, 93
\bibitem[Boyce et al.(2001)]{boy}
Boyce, P.J., Phillips, S., Jones, J.B., et al. 2001, MNRAS, 328, 277
\bibitem[Carretero et al. (2004)]{carre04}
Carretero, C., Vazdekis, A., Beckman, J.E., S\'anchez-Bl\'azquez, P., \& Gorgas, J. 2004, ApJ, 609, L45
\bibitem[Christlein \& Zabludoff(2003)]{chris}
Christlein, D. \& Zabludoff, A. 2003, ApJ, 591, 764
\bibitem[Cole et al.(1994)]{cole}
Cole, S., Aragon-Salamanca A., Frenk, C.S., Navarro, J.F., \& Zepf, S.E. 1994,MNRAS, 271,781
\bibitem[Colless (1989)]{col}
Colless M. 1989, MNRAS, 237, 799
\bibitem[Conselice et al. (2001)]{conse1}
Conselice, C.J., Gallagher, J.S III, \& Wyse, R.F.G. 2001, ApJ, 559, 791
\bibitem[Conselice et al. (2003)]{conse2}
Conselice, C.J., Gallagher, J.S. III, \& Wyse, R.F.G. 2003, AJ, 125, 66
\bibitem[Cortese et al.(2003)]{cor}
Cortese L., Gavazzi, G., Boselli, A., et al. 2003, A\&A, 410, L25
\bibitem[Cross et al.(2004)]{cross}
Cross, N.J.G., Driver, S.P., Liske, J., et al. 2004, MNRAS, 349, 576
\bibitem[Dahlen et al.(2004)]{dahlen}
Dahl\'en, T., Fransson, C., \"Ostlin, G., \& N\"aslund, M. 2004, MNRAS, 350, 253
\bibitem[Davies et al.(2005)]{davies}
Davies, J.I., Roberts, S., \& Sabatini, S. 2005, MNRAS, 356, 794
\bibitem[de Lapparent (2003)]{delapp}
de Lapparent, V. 2003, A\&A, 408, 845
\bibitem[Dolag et al. (2004)]{dol}
Dolag, K., Bartelmann, M., Perrotta, F., et al. 2004, A\&A, 416, 853
\bibitem[Dressler (1978)]{dre}
Dressler, A., 1978, ApJ, 223, 765
\bibitem[Dressler (1980)]{dre80}
Dressler, A., 1980, ApJ, 236, 351
\bibitem[Dressler (2004)]{dre04}
Dressler, A., 2004, in 'Clusters of Galaxies: Probes of Cosmological Structure and Galaxy Evolution', Cambridge University Press, J.S. Mulchaey, A. Dressler, and A. Oemler Jr., eds., p. 206.
\bibitem[Drinkwater et al.(2001)]{drink}
Drinkwater, M.J., Gregg, M.D., \& Colless, M. 2001, ApJ, 548, L139
\bibitem[Driver et al.(1994)]{dri}
Driver, S.P., Phillips, S., Davies, J.I., Morgan, I., \& Disney, M.J. 1994, MNRAS, 268, 393
\bibitem[Durret et al.(2002)]{durr}
Durret, F., Slezak, E., Lieu, R., Dos Santos, S., \& Bonamente, M. 2002, A\&A, 390, 397
\bibitem[Eisenstein et al.(2001)]{eis}
Eisenstein, D.J., Annis, J., Gunn, J.E., et al. 2001, AJ, 122, 2267
\bibitem[Ellingson (2003)]{elli}
Ellingson, E. 2003, Ap\&SS, 285, 9
\bibitem[Feldmeier et al.(2004)]{feld}
Feldmeier, J.J., Ciardullo, R., Jacoby, G.H. \& Durrell, P.R. 2004, ApJ, 615, 196
\bibitem[Fukugita et al.(1995)]{fuk}
Fukugita, M., Shimasaku, K., \& Ichikawa, T. 1995, PASP, 107, 945
\bibitem[Fukugita et al.(1996)]{fuk2}
Fukugita, M., Ichikawa, T., Gunn, J.E., et al. 1996, AJ, 111, 1748
\bibitem[Garilli et al.(1999)]{gar}
Garilli, B., Maccagni, D., \& Andreon, S. 1999, A\&A, 342, 408
\bibitem[Girardi et al.(1993)]{marisa1}
Girardi, M., Biviano, A., Giuricin, G., Mardirossian, F., \& Mezzetti, M. 1993, ApJ, 404, 38
\bibitem[Girardi et al.(1998)]{marisa2}
Girardi, M., Giuricin, G., Mardirossian, F., Mezzetti, M., \& Boschin, W. 1998, ApJ, 505, 74
\bibitem[Goto et al.(2002)]{goto1}
Goto, T., Sekiguchi, M., Nichol, R.C., et al. 2002, AJ, 123, 1807
\bibitem[Graham et al. (2003)]{graham2}
Graham, A.W., Jerjen, H., \& Guzm\'an, R. 2003, AJ, 126, 1787
\bibitem[Gunn \& Gott (1972)]{gunn1}
Gunn, J.E., \& Gott, J.R.III 1972, ApJ, 176, 1
\bibitem[Gunn et al.(1998)]{gunn}
Gunn, J.E., Carr, M., Rockosi, C., et al 1998, AJ, 116, 3040 
\bibitem[Hilker et al.(2003)]{hilk}
Hilker, M., Mieske, S., \& Infante, L. 2003, A\&A, 397, L9
\bibitem[Hogg et al.(2001)]{Hogg}
Hogg, D.W., Finkbeiner, D.P., Schlegel, D.J., \& Gunn, J.E. 2001, AJ, 122, 2129
\bibitem[Horner(2001)]{horner}
Horner, D. 2001, PhD Thesis, University of Maryland
\bibitem[Jerjen et al.(2000)]{jerjen}
Jerjen, H., Kalnajs, A., \& Binggeli, B. 2000, A\&A, 358, 845
\bibitem[Katgert et al.(2004)]{kat04}
Katgert, P., Biviano, A., \& Mazure, A. 2004, ApJ, 600, 657
\bibitem[Kauffmann et al.(1993)]{kau}
Kauffmann, G., White, S.D.M., \& Guiderdoni, B. 1993, MNRAS, 264, 201
\bibitem[Kauffmann et al.(2004)]{kau2}
Kauffmann, G., White, S.D.M., Heckman, T.M., et al. 2004, MNRAS, 353, 713
\bibitem[Liske (2003)]{liske}
Liske, J., Lemon, D. J., Driver, S. P. et al., MNRAS, 2003, 344, 307
\bibitem[Lobo (1997)]{lobo}
Lobo, C., Biviano, A., Durret, F., et al. 1997, A\&A, 317, 385
\bibitem[Loveday (1997)]{lov2}
Loveday, J. 1997, ApJ, 489, 29
\bibitem[Lugger (1986)]{lug1}
Lugger, P.M. 1986, ApJ, 303, 535
\bibitem[Lugger (1989)]{lug}
Lugger, P.M. 1989, ApJ, 343, 572
\bibitem[Lumsden et al.(1997)]{lumsden}
Lumsden, S.L.,  Collins, C.A., Nichol, R.C., Eke, V.R, \& Guzzo, L. 1997, MNRAS, 290, 119
\bibitem[Lupton et al.(1999)]{lup1}
Lupton, R.H., Gunn, J.E., \& Szalay, A.S. 1999, AJ, 118, 1406
\bibitem[Madgwick et al.(2002)]{madg}
Madgwick D.S., Lahav, O., Baldry, I.K., et al. 2002 MNRAS, 333, 133
\bibitem[Marzke et al.(1994)]{marzk}
Marzke, R.O., Geller, M.J., Huchra, J.P., \& Corwin, H.G.Jr. 1994, AJ, 108, 437
\bibitem[Menci et al. (2002)]{Menci}
Menci, N., Cavaliere, A., Fontana, A., Giallongo, E. \& Poli, F. 2002, ApJ, 575, 18
\bibitem[Mercurio et al. (2003)]{Merc}
Mercurio, A., Massarotti, M., Merluzzi, P., Girardi, M., La Barbera, F., \& Busarello, G. 2003, A\&A, 408, 57
\bibitem[Mihos (2004)]{mih04}
Mihos, J.C., 2004, in 'Clusters of Galaxies: Probes of Cosmological Structure and Galaxy Evolution', Cambridge University Press, J.S. Mulchaey, A. Dressler, and A. Oemler Jr., eds., p. 277.
\bibitem[Mobasher et al. (2003)]{Moba}
Mobasher, B., Colless, M., Carter, D. et al. 2003, ApJ, 587, 605
\bibitem[Moore et al.(1996)]{moo}
Moore, B., Katz, N., Lake, G., Dressler, A., \& Oemler, A., Jr. 1996, Nat, 379, 613
\bibitem[Moore et al.(1998)]{moo98}
Moore, B., Lake, G., \& Katz, N. 1998, ApJ, 495, 139
\bibitem[Mulchaey et al.(2003)]{mul}
Mulchaey, J.S., Davis, D.S., Mushotzky, R.F., \& Burstein, D. 2003, ApJS, 145, 39
\bibitem[Murante et al.(2004)]{muran}
Murante, G., Arnaboldi, M., Gerhard, O., et al. 2004, ApJ, 607, L83
\bibitem[Nakamura et al.(2003)]{naka}
Nakamura, O., Fukugita, M., Yasuda, N., et al. 2003, AJ, 125, 1682
\bibitem[Navarro et al.(1996)]{nav1}
Navarro, J.F., Frenk, C.S., \& White, S.D.M. 1996, ApJ, 462, 563
\bibitem[Navarro et al.(1997)]{nav2}
Navarro, J.F., Frenk, C.S., \& White, S.D.M. 1997, ApJ, 490, 493
\bibitem[Paolillo et al.(2001)]{pao}
Paolillo, M., Andreon, S., Longo, G., et al. 2001, A\&A, 367, 59
\bibitem[Phillips \& Driver (1995)]{phidri}
Phillips, S., \& Driver, S. 1995, MNRAS, 274, 832
\bibitem[Phillips et al.(1998)]{phi}
Phillips, S., Driver, S.P., Couch, W.J., \& Smith, R.M. 1998, ApJ, 498, L119
\bibitem[Popesso et al.(2004b)]{pop}
Popesso, P., B\"ohringer, H., Brinkmann J., Voges, W., \& York, D. G. 2004b, A\&A, 423, 449 (paper I)
\bibitem[Popesso et al.(2004a)]{pop1}
Popesso, P., B\"ohringer, H., Romaniello, M., \& Voges, W. 2004a, A\&A accepted, astro-ph/0410011 (paper II)
\bibitem[Popesso et al.(2004c)]{pop2}
Popesso, P., Biviano, A., B\"ohringer, Romaniello, M., \& Voges, W. 2004c, A\&A accepted, astro-ph/0411536 (paper III)
\bibitem[Pracy et al.(2004)]{pra}
Pracy, M.B., De Propris, R., Driver, S.P., Couch, W.J., \& Nulsen, P.E.J. 2004, MNRAS, 352, 1135
\bibitem[Rakos et al.(2001)]{rak}
Rakos, K., Schombert, J., Maitzen, H.M., Prugovecki, S., \& Odell, A. 2001, AJ, 121, 1974
\bibitem[Rauzy et al.(1998)]{rau}
Rauzy, S., Adami, C., \& Mazure, A. 1998, A\&A, 337, 31
\bibitem[Retzlaff(2001)]{retzlaff}
Retzlaff, J. 2001, in 'Clusters of galaxies and the high redshift universe in X-rays. Recent results of XMM-Newtomn and Chandra', XXIst Moriond Astrophysics Meeting, March 10-17, 2001 Savoie, France. Edited by D.M. Neumann  J.T.T. Van.
\bibitem[Sabatini et al.(2003)]{sab}
Sabatini, S., Davies, J., Scaramella, R., et al. 2003, MNRAS, 341, 981
\bibitem[Schechter (1976)]{schechter}
Schechter, P. 1976, ApJ, 203, 297
\bibitem[Schlegel et al.(1998)]{schlegel}
Schlegel, D., Finkbeiner, D.P., \& Davis, M. 1998, ApJ, 500, 525
\bibitem[Smith at al.(1997)]{smi}
Smith, R.M., Driver, S.P., \& Phillips, S. 1997, MNRAS, 287, 415
\bibitem[Smith at al.(2002)]{smith}
Smith, J.A., Tucker, D.L., Kent, S. et al. 2002, AJ, 123, 2121
\bibitem[Stoughton et al.(2002)]{stoughton}
Stoughton, C., Lupton, R.H., Bernardi, M., et al. 2002, AJ, 123, 485
\bibitem[Strateva et al.(2001)]{strateva}
Strateva, I., Ivezi\'c, Z., Knapp, G., et al. 2001 AJ, 122, 1861
\bibitem[Strauss et al.(2002)]{strauss}
Strauss, M. A., Weinberg, D.H., Lupton, R.H., et al. 2002, AJ, 124, 1810
\bibitem[The \& White (1986)]{the}
The, L.S., \& White, S.D.M.  1986, AJ, 92, 1248
\bibitem[Thompson and Gregory (1993)]
Thompson, L.A., \& Gregory, S.A. 1993, AJ, 106, 2197
\bibitem[Trentham (1998)]{tre}
Trentham, N. 1998, MNRAS, 295, 360
\bibitem[Trentham et al. (2001)]{tre01}
Trentham, N., Tully, R. B., \& Verheijen, M.A.W. 2001, MNRAS, 325, 385
\bibitem[Tully et al. (2002)]{tully}
Tully, R.B., Somerville, R.S., Trentham, N., \& Verheijen, M.A. 2002, ApJ, 569, 573
\bibitem[Ulmer et al.(1996)]{Ulmer}
Ulmer, M.P., Bernstein, G.M., Martin, D.R., et al. 1996, AJ, 112, 2517
\bibitem[Valotto et al.(1997)]{val}
Valotto, C., Nicotra, M.A., Muriel, H., \& Lambas, D.G. 1997, ApJ, 479, 90
\bibitem[Van Zee et al.(2004)]{van}
van Zee, L., Skillman, E.D., \& Haynes, M.P. 2004 AJ, 128, 121
\bibitem[Willman et al. (2004)]{will}
Willman, B., Governato, F., Wadsley, J., \& Quinn, T. 2004, MNRAS, 355, 159
\bibitem[Yagi et al.(2002)]{yag}
Yagi, M., Kashikawa, N., Sekiguchi, M., et al. 2002 AJ, 123, 87
\bibitem[York et al.(2000)]{york}
York, D.G., Adelman, J., Anderson, J.E.Jr., et al. 2000, AJ, 120, 1579
\end{thebibliography}
\end{document}